\documentclass{emulateapj}
\pdfoutput=1

\usepackage{amssymb}
\usepackage{amsmath}
\usepackage{algorithm}
\usepackage{algorithmic}
\renewcommand{\algorithmicrequire}{\textbf{Input:}}
\renewcommand{\algorithmicensure}{\textbf{Output:}}

\def\aj{\rm{AJ}}                    
\def\apj{\rm{ApJ}}                 
\def\apjl{\rm{ApJ}}                
\def\apjs{\rm{ApJS}}                       
\def\mnras{\rm{MNRAS}}

\usepackage[usenames,dvipsnames]{color}
\definecolor{purple}{rgb}{0.5,0,0.5}

\newcommand{\algoname}[0]{Zero Bias Regressive Adaptation}
\newcommand{\ac}[0]{ZeBRA}

\newcommand{\algonametwod}[0]{Gaussian Interpolation/Regression Algorithm for Functional Estimation}
\newcommand{\actwod}[0]{GIRAfFE}

\shorttitle{Nonparametric Methods in Astronomy}
\shortauthors{Steinhardt \& Jermyn}
\begin{document}
\title{Nonparametric Methods in Astronomy: Think, Regress, Observe -- Pick Any Three}
\author{Charles L. Steinhardt\altaffilmark{1,2,3,4} \& Adam S. Jermyn\altaffilmark{5}}

\altaffiltext{1}{Cosmic Dawn Center, Niels Bohr Institute, Blegdamsvej 17, 2100 K\o benhavn, Denmark}
\altaffiltext{2}{Dark Cosmology Centre, Niels Bohr Institute, Blegdamsvej 17, 2100 K\o benhavn, Denmark}
\altaffiltext{3}{California Institute of Technology, MC 105-24, 1200 East California Blvd., Pasadena, CA 91125, USA}
\altaffiltext{4}{Infrared Processing and Analysis Center, California Institute of Technology, MC 100-22, 770 South Wilson Ave., Pasadena, CA 91125, USA}
\altaffiltext{5}{Institute of Astronomy, University of Cambridge, Madingley Rd, Cambridge CB3 0HA, UK}

\begin{abstract}
Telescopes are much more expensive than astronomers, so it is essential to minimize required sample sizes by using the most data-efficient statistical methods possible. However, the most commonly used model-independent techniques for finding the relationship between two variables in astronomy are flawed. In the worst case they can lead without warning to subtly yet catastrophically wrong results, and even in the best case they require more data than necessary. Unfortunately, there is no single best technique for nonparametric regression. Instead, we provide a guide for how astronomers can choose the best method for their specific problem and provide a python library with both wrappers for the most useful existing algorithms and implementations of two new algorithms developed here.
\end{abstract}


\section{Introduction}

One of the most common statistical techniques in astronomy is the use of bivariate data to establish and characterize correlations between two properties of a sample of objects.  The goal of regression analysis is to produce a predictor that models the data as closely as possible.  That is, if the data are generated from some underlying function with added noise, the predictor should adhere as closely as possible to the original function.  There are many statistical techniques for performing regression analysis, with the best choice depending upon what is already known about the problem.  Most of the major successes in the recent history of astronomy have relied upon these techniques, perhaps most notably establishing the relationship between recession velocity and luminosity distance \citep{Hubble1929,Riess1998,Perlmutter1999}.  

However, in practice astronomers often select techniques that are poorly suited to their problems and therefore produce suboptimal or even flawed results. Further, because many of these inferior techniques have been widely adopted and their flaws known by statisticians are not well-known within the astronomical community, they are seen as not just acceptable but standard by both colleagues and referees, so that there is limited motivation to choose better techniques that require additional effort and may be met with skepticism.  A major goal of this work is demonstrate why these techniques are flawed and to provide both practical advice on choosing the most suitable regression technique and a library of implementations of the most common methods.

In practice, there are several different related problems that can be thought of as producing a predictor.  For example, there are methods for minimizing noise in estimates of a dependent variable \citep{Stein1956}, interpolation between measurements such as time series data \citep{Hardle1997}, and optimal estimation of the underlying relationship from which the measurements have been drawn.  This last problem is capable of producing significant new astrophysical insights in a purely empirical way (e.g., \citet{Hubble1929}), and is therefore potentially the most valuable.  However, it is also the most prone to difficulties and errors often require a considerable investment of both theory and observation to unravel.  In this work, we therefore focus primarily on recommendations that will be valuable for solving this problem.

The best-studied methods can be used when errors are normally distributed and there is a predetermined model function used to fit to the data.  Perhaps the most common problem, and that encountered by Hubble, is linear regression, seeking to minimize the deviation between the data and a best-fit line.  The best unbiased estimator for normally-distributed errors can be found via ordinary least-squares regression (OLS; cf. \citet{Bevington1969}), with a thorough analysis of other techniques given in \citet{Isobe1990}.  Optimal techniques for linear regression must be modified for use with more complex model functions, although many of the same principles will apply.  

However, often in astronomy observers seek to characterize relationships in which there is no governing theoretical foundation, and thus no specific model function.  These problems can be particularly dangerous because in astronomy uncertainties are commonly poorly modeled, and as a result it can be difficult to determine whether a best-fit function is truly a good description of the data.  In this paper, we consider techniques for nonparametric regression, in which the predictor does not take on a pre-determined form but is rather derived from the data.  Several techniques used commonly in astronomy and other fields are discussed, in addition to new techniques developed in this work. 

The simplest astronomical regression and interpolation problems are those for which errors in both dimensions ($X$ and $Y$) are small compared to the precision with which results are desired.  These problems, which we can think of as having ``0 errors'', merely require interpolation, with methods such as splines most commonly used.  

The one-error (1-E) case, in which the errors in one of $X$ or $Y$ are negligible, is also well-studied.  One-error regressions typically fall into two categories.  Problems with rapidly-changing underlying functions, sparse data, and large errors typically best solved using {\em smoothing} algorithms.  These approximately draw a curve through each individual point, with an attempt to use local information to reduce errors.  The underlying function may have complex behavior in the region between measurements, but insufficient information exists to model that behavior.  Thus, smoothing algorithms typically can only produce a poor fit, and are designed to do so simply and efficiently.  These algorithms are often biased, which is acceptable because the bias will be small compared with the statistical uncertainty.

As more data become available, for any differentiable function the data will look locally linear.  Thus, a better local result can be produced by assuming that linearity to produce a better fit.  The best overall results for such a problem are produced by {\em modeling} algorithms that use not just the data but also priors on the function smoothness and the functional forms that one is likely to encounter in order to incorporate nonlocal information and thereby smooth over the noise without biasing the result.  Modeling algorithms can produce a precise fit, so minimizing bias becomes far more important.

A particularly strong impetus for this work is that although large catalogs and improved measurements are increasingly pushing astronomical problems into the modeling regime, the most common techniques used in the astronomical literature are still variations on a smoothing algorithm.  The common technique of binning data and taking the mean or median value in each bin is in some sense a modeling algorithm.  The model being imposed is that the underlying function is piecewise linear, and that the bin boundaries have been placed at each breakpoint.  However, the reason for using a nonparametric algorithm is that there should be enough information to produce a good model, yet these breakpoints are not known a priori.  Thus, binned averages or medians will be formally invalid.  An alternative, running medians, has a rolling window and is truly a smoothing algorithm.  However, it will produce not just poor but also often biased predictors (\S~\ref{sec:medians}), even when there is enough information to use a unbiased, higher-precision modeling algorithm instead.  Several more statistically robust modeling algorithms already exist, and in this work we also develop a new 1-E technique, \algoname, which uses additional computational resources to present substantial improvements over existing modeling methods.  The goal of this work is to demonstrate why many of the techniques currently used in the literature are dangerous, as well as to provide a python library that makes it easy to use better methods.

The 2-error case, in which both errors in $X$ and $Y$ must be considered, known more formally as ``errors in variables'' \citep{Fuller1987,Fan1993,Carroll2006,Kelly2013} is far more common in astronomy yet also far more difficult to solve.  Indeed, it is common in astronomical literature to see 1-E methods, such as running medians, applied to 2-E cases yielding invalid results.

\begin{figure}[h]
	\plotone{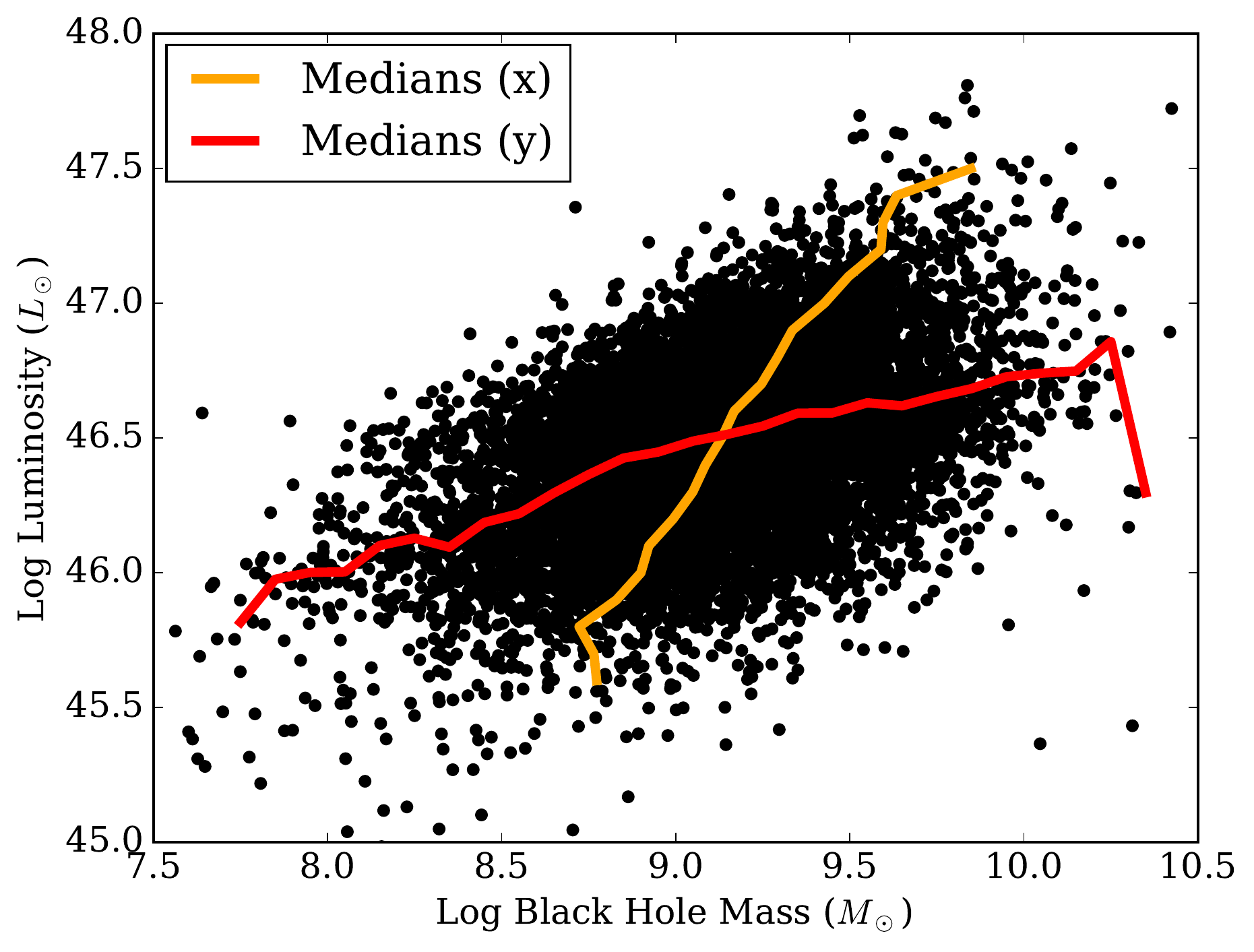}
	\caption{Individual quasars at $1.6 < z < 1.8$ in the mass-luminosity plane, with (log) bolometric luminosity and black hole mass shown from the \citet{Shen2011} catalog.  Running medians were used to calculate predictors $P_X$ (orange) and $P_Y$ (red).  Note that $P_Y$ implies that typical luminosity grows superlinearly with increasing mass, $L \propto M^{1.7}$, while $P_X$ implies that luminosity grows sublinearly with increasing mass, $L \propto M^{0.4}$.  This is a good example of a strong vector symmetry bias (\S~\ref{sec:2d}).}
	\label{fig:qsomeds}
\end{figure}

As an example, consider the relationship between quasar luminosity and mass as measured by the Sloan Digital Sky Survey Seventh Data Release \citep{SDSSDR7}.  For the 105783 quasars in the catalog, we use the redshift, bolometric luminosity and virial black hole mass estimated based upon photometric and spectroscopic properties of each quasar and compiled in \citet{Shen2011}.  Choosing those in the narrow redshift range of $1.6 < z < 1.8$, the quasar mass-luminosity plane (cf. \citet{Steinhardt2010a}) indicates that there is a broad trend towards more massive quasars lying at higher luminosities (Fig. \ref{fig:qsomeds}).  One might hope to use a regression to produce a good description of that relationship.

Running medians (the most common 1-E method) were used to calculate predictors for that relationship.  Binning in $X$ produced an approximately linear predictor $P_X$ in log-log space, indicating that the quasar luminosity $L \propto M^{1.7}$.  Binning in $Y$ also produced an approximately linear predictor $P_Y$, indicating that the quasar luminosity $L \propto M^{0.4}$.  These are very different descriptions of the relationship and would require very different underlying astrophysics to produce them.  Clearly at least one must be incorrect, and in reality both are poor characterizations of the distribution.

It is worth emphasizing that this is not simply a matter of no underlying relationship existing, or of the sampling being poor, or the scatter being too large.  Indeed even for the idealized case of a multivariate gaussian perfectly and infinitely sampled (Fig. \ref{fig:2dgauss}), running medians and other 1-E techniques fail to reconstruct the underlying trend line.  This failure is a direct result of the existence of non-negligible errors in both dimensions, so that solving a 2-E problem with a 1-E method yields a poor result.

In order to compare different techniques, we first describe the most important metrics for determining the quality of a predictor in \S~\ref{sec:metrics}.  By those metrics, running averages and medians perform particularly poorly, as shown in \S~\ref{sec:medians}.  We describe several better, existing methods for regressions in one dimension in \S~\ref{sec:methods}, as well as propose the new \algoname (\ac) technique for nonparametric regression (\S~\ref{subsec:ours}).  The two-error problem, although more common in practice, is less well understood.  A technique for producing an approximate solution with good real-world performance is described in \S~\ref{sec:2d}.  In order to allow a proper comparison between the techniques discussed in this work, we implemented each on a common framework and ran them through a series of tests described in \S~\ref{sec:tests}.  We show that our new techniques often produce the best fit at the cost of the longest runtime, and that if computational resources are a limiting factor, other techniques may be preferable (\S~\ref{sec:results}), though binning never shows up as the preferred option unless the quantity of data is truly enormous and the computational resources are severely constrained.
We discuss options for optimizing these techniques in \S~\ref{sec:discussion}.

The implementations used in this work are made available to the astronomical community in the Supplemental Information and on Github at \href{github.com/adamjermyn}{github.com/adamjermyn}.

\section{Metrics for Evaluating Regression Techniques}
\label{sec:metrics}

The output of a regression performed on $Y$ with respect to $X$ is a predictor $P_X$ which estimates $Y$ given $X$.  Performing the regression with $X$ and $Y$ swapped would yield a representation $P_Y$, estimating $X$ given $Y$.  The key underlying assumption is that these data were produced due to an underlying relationship between $X$ and $Y$, although measurement uncertainties obscure that relationship.  The goal is to produce a predictor $P$ that adheres as closely as possible to the original relationship.  

We will evaluate the performance of different regression algorithms through three primary metrics.  Without loss of generality, here we define these for some underlying $Y(X)$ that we wish to compare with $P_X$.  In each case, the expectation value is taken over the random distribution characterizing the uncertainty in the dataset.

The two most important metrics are the bias and mean squared error (MSE) between the predictor and original function.  The bias, 
\begin{equation}
B \equiv E\left[P_X(X) - Y(X)\right]
\end{equation}
measures the systematic deviation between the predictor and original function.  An unbiased estimator is defined as one with $B = 0$ for all $X$.
It should be noted that this is inherently a distribution- and function--dependent statement, and should not be expected to generalize trivially. For example, ordinary least squares regression is an unbiased estimator of the parameters of a linear relationship for Gaussian noise but not for arbitrary noise distributions and not for arbitrary underlying functions (e.g. if the model is linear but the relationship is non-linear).

The mean squared error (MSE)
\begin{equation}
\mathrm{MSE} \equiv E\left[ \left(P_X(X) - Y(X)\right)^2 \right]
\end{equation}
is an unsigned measure of the typical total deviation between the predictor and the original function.  A related quantity is the root mean squared error (RMSE), given by
\begin{equation}
\mathrm{RMSE} \equiv \sqrt{\mathrm{MSE}}.
\end{equation}
For a model that matches the underlying function and an unbiased estimator, the MSE will be equal to the variance
\begin{equation}
\sigma^2 \equiv \mathrm{MSE} - B^2.
\end{equation}
Indeed in our benchmarks (see section~\ref{sec:results}) we find that the MSE is typically dominated by the variance rather than by the bias for the algorithms considered here.

It is common when evaluating techniques to discuss the so-called bias-variance tradeoff, in which tunable parameters of a model may be adjusted to alter the balance between bias and variance.  This is analogous to the tradeoff between systematic and statistical uncertainties in experimental design.  For many problems, bias is acceptable if the variance will be smaller, because the minimum rms deviation between the predictor and new tests is most important.  However, some problems, including using redshift distributions to constrain cosmology via weak lensing \citep{Laureijs2011,Huterer2006,Masters2015}, instead place their most stringent constraints on the bias.  Unlike experiments, many of the techniques discussed in \S~\ref{sec:methods} have been formally proven to be entirely unbiased.  However, in some cases it may still be correct to choose a biased technique that produces a lower variance.

As with experiments containing both statistical and uncertainties, the overall variance between predictor and function is a combination of an intrinsic scatter and an additional bias.  The best smoothing algorithms reduce the scatter rapidly even for sparse data at the cost of larger bias.  Using these algorithms in a modeling regime will result in a bias-dominated error, so that modeling algorithms with a slightly larger scatter but lower bias will provide a better predictor.  A biased smoothing algorithm will not converge to the original function even in the large-$N$ limit, whereas an unbiased estimator even with large scatter will asymptotically approach the correct answer.  This work is primarily concerned with algorithms suitable for the modeling problems produced by increasingly large astronomical catalogs, and therefore both bias and variance are important metrics.  We only discuss briefly (\S~\ref{subsec:smoothing}) an approach to choosing the appropriate algorithm for smoothing problems.

Finally, in some situations the running time of the algorithm may become important.  Most of the techniques evaluated here choose an approximate solution in order to minimize runtime, and as a result there is a tradeoff between runtime and bias/variance.  Because of the rapid increase in processing capabilities over the past couple of decades, these approximations are now typically only required for the largest datasets or when a large number of regressions must be performed.  We evaluate different algorithms using all three of these metrics.  In practice, the optimal algorithm to use will generally depend on the circumstances and the relative importance of each for that specific use case.

\subsection{Symmetry Bias}

An additional property which should be considered is symmetry between calculating $Y(X)$ and $X(Y)$.  Clearly when attempting to describe the underlying relation, producing $X(Y)$ given $Y(X)$ should be a matter of algebra rather than performing an additional regression, so long as both functions are monotonic.  However, as with the quasar dataset shown earlier (Fig. \ref{fig:qsomeds}), with some techniques in practice $P_X$ and $P_Y$ may be very different functions.

This was discussed in detail by \citet{Isobe1990}, with the conclusion that many standard regression techniques fail to be symmetric.  There are two associated metrics, the vector symmetry bias,
\begin{equation}
B_S \equiv \left(E\left[P_Y(P_X(X)) - X\right],E\left[P_X(P_Y(Y))- Y\right]\right)
\end{equation}
and the scalar symmetry variance
\begin{equation}
\sigma^2_S \equiv E\left[\left(P_X(X) - Y\right)^2\right] + E\left[\left(P_Y(Y) - X\right)^2\right].
\end{equation}
Intuitively, large symmetric bias corresponds to systematically having differences between the estimates made in $X$ and $Y$, while large symmetric variance corresponds to having the models make large unsigned symmetry errors.  

The principal reason that symmetry bias will arise is that errors in the two variables are being considered asymmetrically with respect to each other.  In the case of one-error techniques, this means assuming that all of the error is contained in one variable.

\begin{figure}[h]
	\plotone{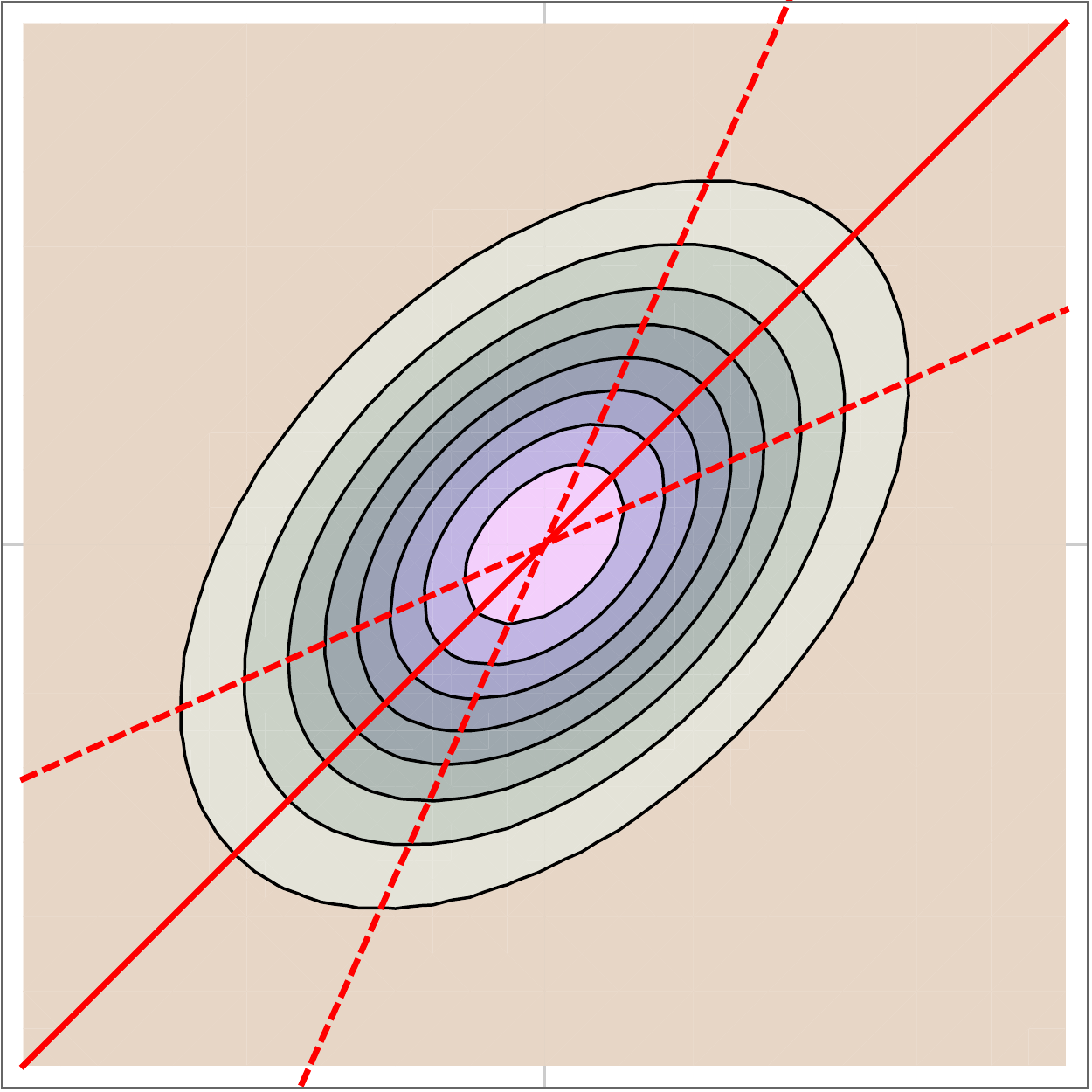}
	\caption{Running medians in the limit of an infinite number of perfectly-measured points drawn from an underlying Gaussian distribution in two orthogonal, but off-axis dimensions.  There is a strong symmetry bias, because the locus traced out by the median $X$ given a particular $Y$ is indeed not the same as that from the median $Y$ given $X$.  The solid line indicates the correct fit, which is given by neither 1-E interpolation.}
	\label{fig:2dgauss}
\end{figure}

An example where this fails badly for a two-error problem is shown in Fig. \ref{fig:2dgauss}.  In this case there is a true, underlying relationship between the two variables $X$ and $Y$ given by the solid line.  Such relationships in astronomy are typically driven by stochastic feedback mechanisms, so that the observed relationship is convolved with a combination of intrinsic scatter and measurement error, both modeled as normally distributed zero-mean distributions.  For the methods discussed in this work, the distinguishing between intrinsic (physical) scatter and measurement error is typically not of primary interest, so we treat the combination of the two as a combined error in measuring the underlying relationship.

In astronomy it is this relationship that is usually of interest because this is what physical laws constrain.  Estimates of $Y$ given $X$ and of $X$ given $Y$ are predictive in the sense of producing unbiased estimands, but they are poor approximations of the underlying relationship between the two variables, and so using predictive estimands as a substitute for this is a fundamental error.  Thus, the solid line in Fig. \ref{fig:2dgauss} is the desired fit, even though both dashed lines are valid solutions to different problems.

Whether symmetry bias actually indicates a problem depends upon the nature of the measurement uncertainties.  For a 1-E regression, this is not a problem, and indeed the asymmetry reflects the asymmetry of the errors.  Techniques which assume one variable is measured perfectly and the other is not will only be valid with the correct choice of independent and dependent variables, so we should not expect symmetry.  Thus, we do not consider symmetry bias when evaluating the 1-E techniques in \S~\ref{sec:results}.  In most astronomical scenarios, however, the uncertainties are in both quantities, and therefore it is important to pick a technique that does not assume otherwise~\citep{Stefanski1985, Stefanski1987, Carroll2006}.  As result, we suggest that this property should be considered more carefully in evaluating two-error regression techniques, since a symmetry bias is a good indicator that a poor fit has been produced.

\section{Averages and Medians}
\label{sec:medians}

The most commonly used techniques for nonparametric regression in astronomy are variations on binned averages and binned medians, both of which are smoothing tools.
One variable is selected as the independent variable (without loss of generality, $X$).
Objects are then binned in $X$, with the mean or median $Y$ calculated in each bin.
Connecting those with linear segments produces a predictor $P_X$ for the underlying function.
A common variation is running averages or medians, in which a different bin is calculated centered around each point.
Because medians are less sensitive to extreme outliers, they are more commonly selected in the astronomical literature.

As we describe in the remainder of this section, binned averages and medians are deeply flawed even as smoothing routines and therefore should be avoided.
Running averages and medians are better choices when the problem is suitable for a smoothing algorithm, but are often incorrectly applied to problems that should be modeled instead (\S~\ref{subsec:smoothvsmodel}).
For a sufficiently simple function and well-measured data, however, even deeply flawed interpolation routines will perform decently.
Further, in smoothing applications errors from poor interpolation are generally smaller than uncertainties resulting from measurement errors or small samples.
As a result, most of the published work relying upon binned and running medians for smoothing has likely drawn correct conclusions.

At a minimum however, flawed statistical methods will generally lead to an overestimate of the amount of data required to draw a conclusion, and hence can lead to a mis-allocation of telescope time.  Even more worryingly, there also exist astronomical problems in which running medians would suggest an entirely wrong conclusion in which there are no a priori warning flags.  For example, running medians will indeed produce a predictor given the quasar dataset shown earlier (Fig. \ref{fig:qsomeds}), but that predictor gives a misleading view of the underlying relationship because it has applied a 1-E technique to a 2-E problem.

\begin{figure}[h]
	\plotone{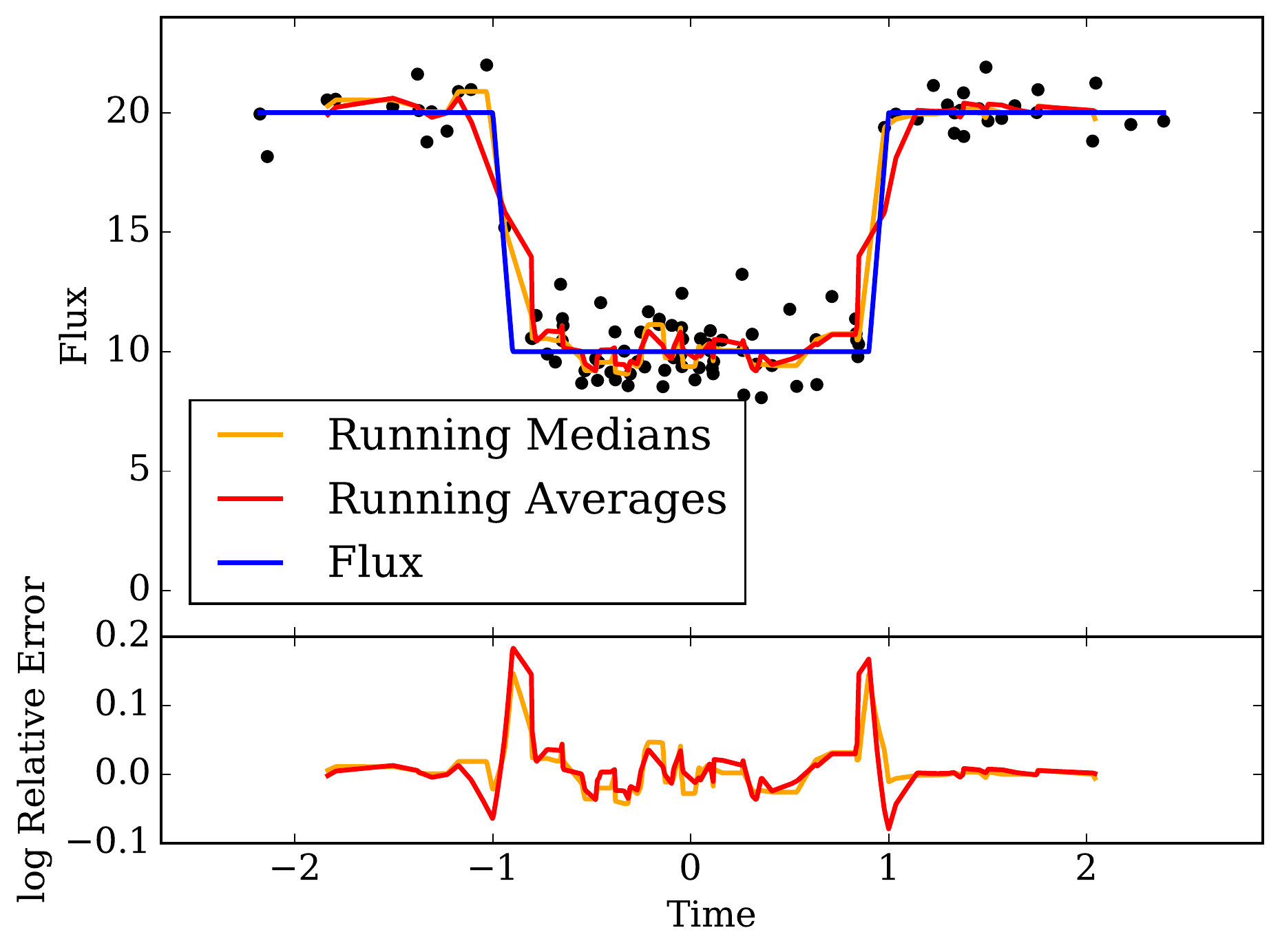}
	\caption{Running average (red) and running median (orange) regressions for 100 points (black) selected normally in $X$ from a model exoplanet light curve (blue). Each running window is five points wide.  These techniques are biased where the distribution is curved, making the eclipse appear systematically more gradual because they are designed as smoothing algorithms.  Without a correction, this would lead to systematic overestimation of the orbital inclination by exaggerating the duration of the flux decline and reducing the duration of the flux minimum.}
	\label{fig:lightcurve}
\end{figure}

Even for a problem that is indeed one-error, running averages and medians can produce flawed results.
They were designed not as fitting algorithms but as smoothing algorithms.
Thus, fitting a sharply-changing function running medians will produce a more gradual change.
If used improperly this might, for example, cause one to systematically overestimate the orbital inclination of a transiting exoplanet from its light curve (\ref{fig:lightcurve})

\footnote{It should be noted that exoplanet transits are properly described using parametric rather than nonparametric regression techniques, so that running medians should never be used for such a task.}.

Unfortunately, whether a distribution is pathological when modeled by a smoothing method depends upon properties of the underlying function rather than on properties of the dataset.
As a result, it can be difficult to determine by inspection whether using medians has mis-characterized a distribution.
This is of particular concern because the goal of many algorithms is to be resilient even in a worst-case scenario \citep[cf.][]{Pease1980}, at least in terms of the underlying relationship.
We describe more formally some of the key reasons that running medians and related methods are flawed below.

\subsection{Binning is Dangerous}

The first step in producing means or medians, binning data, is an inherently dangerous proposition. There are two reasons for this. First, binning `fails silently' and so may be readily applied blindly, often with disastrous results. Secondly, there are intrinsic problems with binning as a regression technique. Here we address the first concern. We examine the second one at various points later in this section.

Inherently, binning data properly requires that three things all be true:
\begin{enumerate}
\item{Points placed in the same bin truly represent the same quantity.}
\item{Points placed in different bins truly represent different quantities.}
\item{There are no hidden variables improperly accounted for in choosing bins.}
\end{enumerate}

The second of these is the weakest constraint, as a failure to obey it will merely lead to drawing results from fewer data points than possible, and thus a suboptimal result.  The other two can be more dangerous and in extreme cases may lead to wildly misleading results.

\begin{figure}[h]
\includegraphics[width=0.5\textwidth]{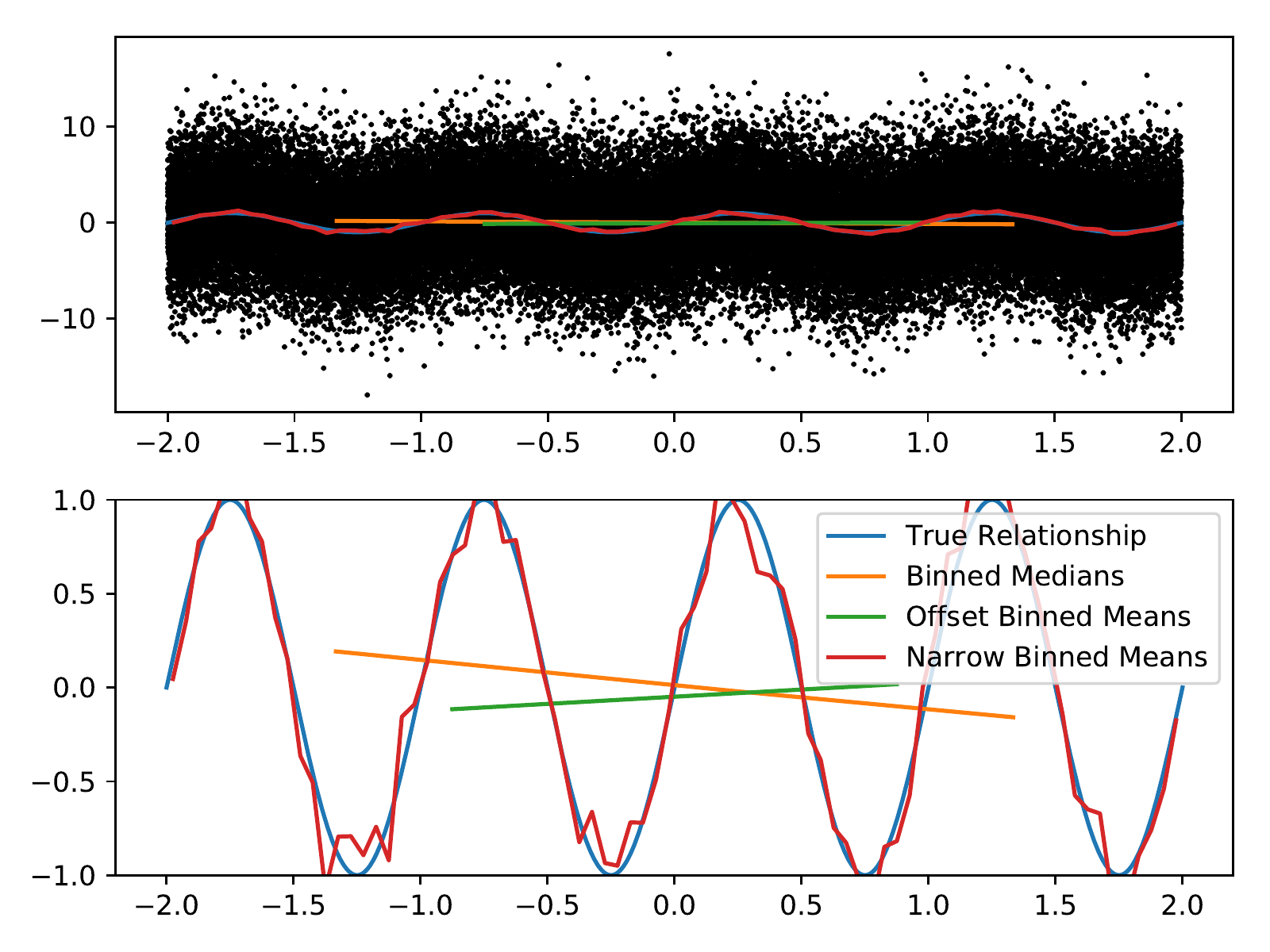}
\caption{An example dataset is shown which was generated by adding normally distributed noise of amplitude $4$ to a unit amplitude sinusoid. The top panel shows the data along with three non-parametric fits and the underlying sinusoid, while the bottom shows just the fits and sinusoid in a zoomed view. Despite the sinusoid being clearly visible in the point cloud, two instances of binned medians (using bin widths offset from the period) yield opposing trends (green and orange) depending on where the edge of the first bin is placed. By contrast binned means with a smaller width yield a good approximation of the underlying trend.}
\label{fig:oscillations}
\end{figure}

As an example of problems when the first constraint does not hold, consider a dataset with 64000 points generated from several periods of a sinusoidal function with noise (Fig. \ref{fig:oscillations}).  When placed into bins 5\% of a period, averaging produces a reasonable estimate for the underlying function.  However, when placed into broader bins of width 1.1 periods, the sinusoidal behavior is lost and the function appears to be linear.  Further, depending upon the exact offset of these broad bins, the function might appear to be either monotonically increasing or monotonically decreasing.  This is a result of choosing the wrong bin size, so that points which should be considered different because of the underlying function are mistakenly placed in the same bin. Unfortunately, choosing a safe bin size does often require knowledge about the underlying function, which makes binning dangerous as a poor nonparametric technique.

When the final constraint on binning, avoiding hidden variables, does not hold the result is a generic problem that can apply not just to binned methods but more generally to regressions when not all variables are accounted for.  Partitioned data can often show different trends than aggregate data due to the presence of hidden variables \citep{Simpson1951}, an effect often termed Simpson's Paradox or the Yule-Simpson effect.  As a constructed example to illustrate this effect, imagine a survey measuring the extinction of the stellar population in both star-forming and passive galaxies at fixed mass and at two different redshifts, yielding the measurements given in Table \ref{tab:simpson}.  Most galaxies at this mass are star-forming at high redshift but passive at low redshift, thus the sample is mostly star-forming at high-$z$ but mostly quiescent at low-$z$.
\begin{center}
\begin{table}
\caption{Fictional data on stellar population ages in star-forming and passive galaxies, constructed to illustrate Simpson's Paradox.}
\label{tab:simpson}
\centering
\begin{tabular}{ |c|c|c|c| } 
\hline
$z$ & $N$ & Type & Avg. E(B-V) (mag.) \\
 \hline
3 & 900 & SF & 0.20 \\
3 & 100 & Passive & 0.10 \\ 
0.2 & 100 & SF & 0.60 \\
0.2 & 900 & Passive & 0.30 \\
\hline
All & 1000 & SF & 0.24 \\
All & 1000 & Passive & 0.28 \\
\hline
\end{tabular}
\end{table}
\end{center}
At either fixed redshift, this survey would conclude that star-forming galaxies are dustier than quiescent ones.  However, grouped only by activity and not by redshift, instead the average passive galaxy in this survey has more extinction than the average star-forming one.  In this case, the hidden variable is redshift: an analysis ignoring the strong trend towards higher E(B-V) with lower redshift can draw a poor conclusion.

Simpson's Paradox is not restricted to frequencies, but can also be present in measuring correlations.  An example of a problem in which this effect has been important in astronomy is studies of the influence of position within a galaxy cluster on AGN activity.  Different studies have concluded that there is a higher frequency of AGN towards the centers of clusters \citep{Ruderman2005}, a higher frequency towards cluster outskirts \citep{Gilmour2007}, or a constant frequency \citep{Miller2003,Sorrentino2006}.  For these studies, selection effects make it very difficult to produce a control sample that isolates the effect of radius from the center of the cluster and marginalizes over the many other important variables \citep{Martini2007,Khabiboulline2014}.

Simpson's paradox is somewhat of a misnomer because there is no actual paradox involved.  Incorrect assumptions about a lack of hidden variables and incomplete information will lead to seemingly contradictory conclusions.  However, a correct understanding of all of the important variables involved will lead to a consistent result.   In our fictitious example, it is indeed a fact that the average observed passive galaxy has a higher extinction than the average observed star-forming galaxy.  However, this is likely not the question that most astronomers are interested in answering.

An important principle in developing programming syntax and coding conventions is that wrong code should look wrong.  It is similarly ideal that wrong statistical analysis should look as obviously flawed as possible, and techniques with this property are more practical choices.  The most dangerous aspect of binning is that it is guaranteed to yield an answer regardless of whether there are important hidden variables.  In most cases, it is difficult to determine a priori whether different binning and selection criteria would yield a different result.  

Even in cases with no explicit hidden variables, measurement uncertainty will result in some data being placed in the `wrong' bin.  Even for some of the simplest cases, such as a using running medians on a bivariate sample to approximate an underlying linear relationship, binning can induce biases if sample selection is not uniform in the binned variable.  The specific assumption made in binned averages or medians is that the underlying function is linear and consistently measured within the domain included in each bin, but this stops being true at the bin boundary.  Without prior knowledge of the underlying function, this assumption will almost certainly be incorrect, and if there is prior knowledge, parametric techniques will produce a superior fit.

In short, methods involving binning are dangerous, are very often flawed, and should be reserved for the rare class of problems that are so well understood that one can be confident every important variable is properly accounted for.  For most problems in astronomy, binning should be avoided as part of properly conservative data analysis.

\subsection{Bias}

One of the main reasons to use averages or medians is that they are computationally efficient even for extremely large samples.   However, for most practical applications, even running bins will produce a biased predictor.  That is, on average over a large number of trials, there will be a systematic offset between the predictor and the underlying function as shown below.  The most common variation of these techniques uses a fixed-width bin in the dependent variable for further simplicity of implementation.  In that case, even in the limit as the number of measurements increases without bound, a running averages predictor will not converge to the underlying function.  

\subsubsection{Averages}

The more biased of the two techniques is averages.  For a linear underlying function, it is indeed true that $y(x)$ can be estimated by averaging $y$ calculated at nearby values of $x$.  However, for a nonlinear $y(x)$, this assumption will break down.  Assuming $y(x)$ is differentiable, running averages will approach the correct value as the bin width goes to 0.  However, for finite bin width, even running averages will produce a biased approximation to a nonlinear function, and generally the bias will increase as the bin size becomes larger.

To illustrate, consider perfect measurements of pairs $(x,y)$ lying on $y = x^2$ over the domain $x > 0$.  We wish to create the predictor $Y(X)$ using a bin width of $2\Delta x$ and use it to predict $Y(x_0)$, where we will consider $x_0 > \Delta x$.  Let points ${x_i}$ be uniformly selected in $x$ within the bin $[x_0-\Delta x,x_0+\Delta x]$.  Then, the predictor $Y(X)$ will be the average $y(x_0)$ over that distribution, and thus on expectation,
\begin{equation} 
E[Y(x_0)] = \frac{\int_{x_0-\Delta x}^{x_0+\Delta x}{x^2 dx}}{2x} = x_0^2 + \frac{{\Delta x}^2}{3}.
\end{equation} 
Thus, the bias, $B \equiv E\left[P_X(X) - Y(X)\right] = {\Delta x}^2/3,$ is nonzero and remains fixed at constant bin size even as the number of perfect measurements $N \rightarrow \infty$.  It is possible to avoid this by having bin sizes shrink as $N\rightarrow \infty$, but the bias may never be completely eliminated in this way, resulting in the need to gather more data to achieve the same precision when compared with unbiased methods.

\subsubsection{Medians}

At first glance, medians appear more robust than averages.  For example, consider a predictor for $y=x^2$ as described above.  Because $y(x)$ is monotonic, ordering the points within a bin in $x$ produces an identical ordering to doing so in $y$.  Thus, selecting the median $y$ in this bin will select $x^2$ for the median $x$ in the bin.  For any perfectly measured monotonic function, medians will always select points from the sample, and thus unlike averages will always select points lying on the original function.  However, for a finite sample size, we show even running medians will produce a biased estimator for similar reasons to running averages.  

Selecting $x$ uniforming from over $[x_0-\Delta x,x_0+\Delta x]$, the probability density is 
\begin{equation}
p(x)dx = \left\{
     \begin{array}{@{}l@{\thinspace}l}
       \frac{dx}{2\Delta x}, & x \in [x_0-\Delta x,x_0+\Delta x] \\
       0, & \mathrm{otherwise} \\
     \end{array}
   \right.
\end{equation}
Thus, for $x$ selected from a uniform distribution within the bin, then, on expectation the median value of $x$ will be the center of the bin $x_0$ for any sample size.  

The probability density function for $y$, then, will be \begin{equation}
p(y)dy = \left\{
     \begin{array}{@{}l@{\thinspace}l}
       \frac{dx}{2\Delta x}, & x \in [x_0-\Delta x,x_0+\Delta x] \\
       0, & \mathrm{otherwise} \\
     \end{array}
   \right.
\end{equation}
The median value in this distribution satisfies
\begin{equation}
\int_{(x_0-\Delta x)^2}^{y_{\mathrm{median}}} \frac{dy}{4\Delta x \sqrt{y}} = \frac{1}{2\Delta x}\left(\sqrt{y_\mathrm{median}} - x_0 + \Delta x\right) = \frac{1}{2}.
\end{equation}
Solving indeed yields
\begin{equation}
y_\mathrm{median} = x_0^2.
\end{equation}
Thus, as the sample size $N$ increases, the number of bins of fixed size increases and hence the bin width decreases.
In this limit then running medians will converge to the correct answer,
\begin{equation}
\lim_{N \rightarrow \infty} P_x(x_0) = x_0^2.
\end{equation}

For finite $N$, however, $P_X$ will be a biased estimator.  For a specific trial $t$, the median $x_t$ within the bin will not be exactly $x_0$, but will be drawn from a symmetric distribution about $x_0$.  The predictor $P_{X,t} = y(x_t) = x_t^2$.  
$E[x_t] = x_0$, by symmetry.  Averaging over a symmetric distribution of $x_t$ will produce $E[x_t^2] > x_0^2$, as for running averages above.  For $N = 1$, running medians are exactly identical to running averages.  As $N$ increases and the distribution of $x_t$ becomes narrower around $x_0$, the bias will decrease.  In the limit as $N \rightarrow \infty$, the distribution of the median value in the bin becomes a Kronecker delta function $\delta(x_0)$, and thus $\lim_{N \rightarrow \infty} P_X = x_0^2$.  A full treatment for finite $N$ is calculable but beyond the scope of this paper, although several specific values are given in \S~\ref{sec:tests}.

Running medians will produce a better result than running averages, particularly for large $N$.  However, both are biased for nonlinear functions.

\subsection{Interpolation Error: Smoothing vs. Modeling}
\label{subsec:smoothvsmodel}

Beyond the bias discussed in the previous two sections, there is also interpolation error associated with evaluating functions in-between binned results.
To understand how this compares with noise-based bias, suppose that we have imperfect measurements and connect the dots between neighboring points.
Then the mean squared error in a curve between a point at $x_0$ and a point at $x_1$ with unbiased noise distribution $\eta(x)$ is
\begin{align}
	E =& \frac{1}{x_1-x_0}\int_{x_0}^{x_1}\int_{-\infty}^{\infty}\int_{-\infty}^{\infty} \left[f(x) - \frac{x - x_0}{x_1 - x_0}\left(f(x_0) + \eta(x')\right)\right.\\
	&\left. - \frac{x_1 - x}{x_1 - x_0}\left(f(x_1) + \eta(x'')\right)\right]^2 dx' dx'' dx\nonumber\\
	\approx & \frac{1}{4} \Delta x^4 f''(x)^2 + \frac{2}{3}E[\eta^2].
\end{align}
Note that the bias in the same window will be non-zero, and given by the curvature of the function, as the noise will average to zero.

Now if our point values were determined by a binned estimator with $N$ points, the noise portion of the variance would fall by a factor scaling as $\sqrt{N}$, though there might still be bias unaccounted for as a result of the finite bin width.
This means that the three sources of error in our estimators are:
\begin{enumerate}
	\item Noise-derived variance
	\item Bin-derived bias
	\item Interpolation bias and variance
\end{enumerate}
If we avoid the use of binned methods or use large enough $N$ that bin bias is small, the remaining two sources of error determine whether the problem is one of smoothing or modeling.
In particular, if the interpolation is order $k$ and
\begin{equation}
	\frac{(\Delta x)^{2(k+1)}}{(k+1)!}\left|\frac{\partial^k f}{\partial x^k}\right|^2 \ll E[\eta^2],
	\label{eq:decision}
\end{equation}
then the problem is a modeling problem, as the interpolation error is small but the function will be difficult to recover in any meaningful way from the noise unless non-local information and meaningful priors on noise distribution are used.
In the opposing limit interpolation error dominates and without strong priors on the function space very little can be determined accurately.
This is the smoothing regime.
Note that we have omitted a factor of $1/\sqrt{N}$ from the right hand side in Eq.\ \eqref{eq:decision}.
This is intentional, as $N$ is derived either from the quantity of data or from a prior on the bandwidth of the function, and in either case once $N$ is specified the problem is definitely a smoothing one.

The dependence of Eq.\ \eqref{eq:decision} on $k$ suggests that it is possible to switch between the modeling and smoothing regime just by modifying the order of the interpolation method used.
This is a useful option, but it is worth being careful when employing it.
Increasing $k$ reduces the interpolation error and makes the problem more solidly a modeling one, but at the cost of introducing additional free parameters which reduce the statistical significance and the likelihood of overfitting.
Decreasing $k$ increases the interpolation error, making the problem appear more like a smoothing one, but at the cost of potentially wasting the opportunity to extract more useful information from the data.
Thus the choice of $k$ should actually be motivated by questions of statistical significance rather than by an attempt to alter the class of the problem.

\subsection{Choosing a Smoothing Algorithm}
\label{subsec:smoothing}

Although the remainder of this work is focused on modeling problems, we briefly discuss the difference between two common smoothing techniques.  Medians are far more common in the astronomical literature, but averaging is also appropriate for some smoothing problems.  Formally, medians and means for a constant function are optimal estimators under the L1 ($\chi$) and L2 ($\chi^2$) loss functions, respectively.  To see this, consider a sample of size $N$ of a random variable $x$.

The L1 norm of the error in an estimator $\beta$ is
\begin{equation}
	E = \frac{1}{N}\sum_i |\beta - x_i|.
\end{equation}
Differentiating with respect to $\beta$ gives
\begin{equation}
	\frac{dE}{d\beta} = \frac{1}{N}\sum_i \text{sgn}(\beta-x_i).
\end{equation}
This is zero when half of the samples lie above $\beta$ and half lie below.  In the case where there is an odd number of samples the root occurs when $\beta$ lies on the median point.  Thus the median of the sample satisfies the criteria of minimizing the L1 norm.

The L2 norm of the error in $\beta$ is
\begin{equation}
	E = \frac{1}{N}\sum_i \left|\beta - x_i\right|^2 = \frac{1}{N}\sum_i (\beta - x_i)^2..
\end{equation}
Differentiating with respect to $\beta$ and extremizing gives
\begin{equation}
	\beta = \frac{1}{N}\sum_i x_i,
\end{equation}
which is the mean.

Thus, the proper choice of smoothing algorithm depends upon the nature of the errors on individual measurements.  If the errors are dominated by Gaussian noise, then minimizing $\chi^2$ is correct, so averaging will be the best choice.  This is even the case if only the relative errors are known, as minimizing $k \chi^2$ produces the same answer as minimizing $\chi^2$ for constant $k \neq 0$.  For example, there may be an underlying function that connects two quantities on average, a Gaussian measurement error in one of them, and an additional physical scatter in the relationship.  Both scatter measurements off the average underlying relationship, producing an error function that is the convolution of the two.  If that convolution is approximately Gaussian, then averaging will be the best choice.  

Medians are preferred when the error distribution has larger tails than a Gaussian distribution, as then an L1 norm is less sensitive to extreme outliers that an L2 norm will model as features.  For example, if a sample includes a handful of mis-categorized objects that lie on a much different relation, medians will be more effective at approximating the original function.  More generally, the L1 norm prefers 'compact' descriptions of data, which results in an estimate which is more likely to disregard points as noise\ \citep{Tibshirani94regressionshrinkage}.

\subsection{Prospects}

Although for many distributions running averages and medians will yield an acceptable result, they are formally biased.  Perhaps more importantly, they are non-trivially biased in many practical cases, even for high-quality data.  This bias is introduced both when choosing bins and when taking means or medians with a bin.  Further, it is often difficult to determine with certainty that a distribution is non-pathological.  That is, wrong results do not always look wrong.  Although lack of symmetry when fitting both possibilities of dependent and independent variables can indicate that running medians have yielded a poor result, not all symmetric fits are correct.  The only proper measures of bias and variance are bias and variance.

\begin{figure}[h]
\plotone{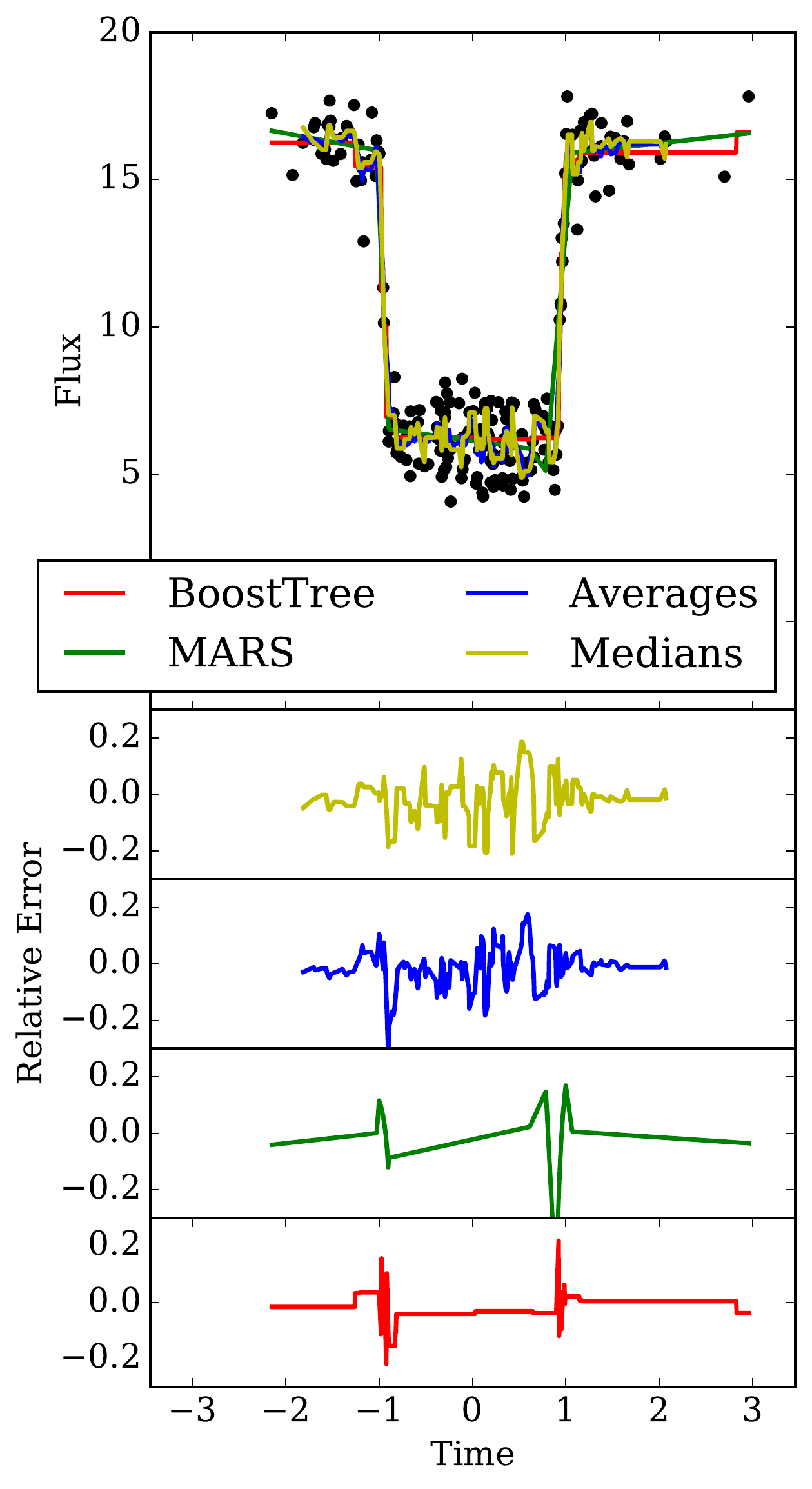}
\caption{Regression on 200 points (black) selected randomly from a uniform distribution in $X$ from a model exoplanet light curve fit using four techniques: running averages (blue), medians (gold), Boosted Trees (red), and MARS (green), the same test shown for running averages and medians in Fig. \ref{fig:lightcurve}.  The latter two algorithms are described in \S~\ref{sec:methods}, and they perform far better in approximating the underlying function, particularly in determining the depth of the eclipse and steepness of its walls, which are used to determine ratio of the planetary radius to the stellar radius and the inclination respectively.}
\label{fig:lightcurve2}
\end{figure}

By those metrics, running averages and medians perform reasonably as smoothing algorithms, in that they quickly produce a low variance for small datasets.  However, they are very poor modeling algorithms, both theoretically in the sense that they are biased and in comparison with other algorithms on large datasets.  Indeed, the alternative methods discussed in \S~\ref{sec:methods} all produce reasonable agreement and a much better description of the underlying trend for the exoplanet transit light curve (Fig. \ref{fig:lightcurve2}) fit with running averages and medians earlier.  The dataset is not inherently pathological, but rather running medians produce a poor result when compared with the more advanced modeling techniques discussed in the following section.  

A similar result can be produced even when considering a linear underlying relation.  Even though running averages and medians will now be unbiased, they are still higher-variance estimators than the alternatives (\S~\ref{sec:results}) for even intermediate-size datasets.  The proper conclusion is that when the underlying function is not well-understood, as is typical when a nonparametric regression is the analytical tool of choice, conservative analysis is best done using tools that are not dependent on binning and which are known to suffer from fewer of these pitfalls.

\section{Nonparametric Regression Techniques for Modeling Problems}
\label{sec:methods}

Having concluded that running averages and medians are often unacceptable despite their ubiquity, here we summarize many existing statistical techniques for nonparametric regression that are more suitable for modeling problems as well as introduce a new algorithm.  We have benchmarked these tools in \S~\ref{sec:tests}.

In principle, there are many more ways to pick a suboptimal algorithm than considered in this work.  For example, a binned algorithm with a fixed number of bins will not asymptotically converge to the correct answer as the number of points increases.  Alternatively, selecting a 1-E algorithm for a 2-E problem may asymptotically converge, but to an answer that is not indicative of the underlying relationship.  It is hoped that the discussion in this work will already convince the reader that such methods are unsuitable under essentially all circumstances.  As a result, the methods discussed below and benchmarked in \S~\ref{sec:tests} are chosen to all be reasonable choices that asymptotically converge. However, depending upon the specific problem being solved, some will converge far more efficiently than others.

\subsection{Local Polynomial Regression}

Local polynomial regression consists of fitting a polynomial in a running window and so is a special case of LOESS/LOWESS regression~\citep{9781412975148}.
Often the data are weighted with a kernel that moves alongside the fit to ensure a smooth and well-behaved prediction.  Local regression with arbitrary-order polynomials is available off-the-shelf via the PyQT-fit Python package.

A closely related and perhaps better-known method is that of smoothing splines.
These use the data to place knots and then fit spline functions to those.
This has the advantage of producing an analytic function which may be efficiently stored and differentiated~\citep{9780412300400}.  Smoothing splines of various orders are supported by the SciPy Python package.

\subsection{Multivariate Adaptive Regressive Splines}

Multivariate Adaptive Regression Splines (MARS) is a semi-parametric regression method developed by \citet{Friedman1991} to fit data in several dimensions.  Spline interpolation defines the predictor piecewise with functions of the form
\begin{equation}
f(x) = \max\left(0,x-a\right),
\end{equation}
where $a$ is a constant, requiring that the pieces connect with a high degree of smoothness.  The most common implementations of spline interpolation (cf. \citet{Press1986}) use cubic B-splines to produce a smooth, 4th-order function.  MARS improves upon this by using a variety of common functions as part of its basis set, producing a predictor as a combination of those basis functions.  Thus, different functional forms may be used over different piecewise domains in order to compose the final predictor.  As part of its optimization, MARS builds an overfit function on an initial pass, then prunes its fit by removing the least effective term until termination.

When the underlying function can indeed be perfectly decomposed in terms of the given basis functions, MARS will be unbiased.  For other underlying functions, MARS can exhibit bias for certain distributions when the basis functions cannot provide a good approximation to the underlying form of the data.  In practice, many functions are well-represented by the basis functions in the standard MARS suite of splines, and therefore it performs with extremely low bias.  
MARS is a trademarked and licensed technique, although an off-the-shelf open source implementation of this technique is available in Python through the Py-Earth package.  The authors had difficulty with the various dependencies, and were unable to run the Python 3.5 version.  Benchmarking on MARS was done using Python 2.7.  Although in principle it is possible to run MARS with a custom set of basis functions, doing so can be difficult in practice without a full understanding of the algorithm and its implementation. 

\subsection{Boosted Trees and Random Forests}

A relatively recent advance in regression has come from ensemble techniques, which combine many prediction models to produce a strong unbiased predictor. Two powerful examples of ensemble methods are Gradient Boosting and Bagging.

Gradient Boosting combines many weak prediction models which typically have high bias to produce a strong unbiased predictor~\citep{Friedman2001}. This is commonly implemented with decision trees in the form of Boosted Tree regression, also known as Gradient Boosted Trees.
This technique is related to MARS in that it amounts to building a function from an ensemble of different functions fit to the data.
Central to the idea of Boosted Trees is that additional model complexity is introduced to reduce the bias until the variance is low enough that overfitting appears likely.
This may be implemented in practice with cross-validation, wherein the quantity being optimized is not the overall fit but rather the likelihood that a fit of the data with one point missing accurately reproduces that point.
The details of this are actually not all that important, as gradient boosting is generally quite robust to overfitting by virtue of it being formed of a large number of poor estimators, each of which is given low weight.

In contrast to Gradient Boosting, Bagging \citep{Breiman2001} is typically used with several strong predictors. As with Boosted Trees, these are usually implemented with decision trees in the form of Random Forest regression. The key difference is that while Gradient Boosting iteratively fits weak predictors to form more complex models, Bagging produces a weighted average over strong predictors. A key advantage of Bagging over Gradient Boosting is that it is much more readily parallelized.

It is worth emphasizing that these ensemble methods are not guaranteed to be unbiased for arbitrary data, although they may be implemented in such a way as to produce a statistically optimal tradeoff between bias and variance, contingent upon some prior over the underlying parameter space. In practice, because these methods are highly flexible, it is important to be aware of the choices made by a given implementation, as this will impact the quality of the fit and the extent to which this method is appropriate for a given dataset.

Off-the-shelf open source implementations of both Boosted Trees and Random Forests with various different bias-variance tradeoff options are available through the Scikit-learn Python package.
In this work we have used Boosted Trees implementation with a least squares loss function, $100$ estimators, a learning rate of $0.1$, corresponding to the significance assigned to individual estimators, and a tree depth of $1$. We have also used Random Forests with $10$ estimators and at least $2$ samples per split.

\subsection{Kriging}

Kriging, or Gaussian Process Regression, is an interpolation technique which models the underlying function as a Gaussian random process \citep{Wahba1990,Williams1997}.  It is designed for a situation in which a dataset is sparse but well-measured, and optimized to predict intermediate values in regions with no data.  If these assumptions are correct, Kriging is not only unbiased but also produces the minimum variance linear unbiased estimator of these intermediate values.

The problem of finding a minimum variance unbiased estimator does not have a general solution and as a result the assumptions made in Kriging will not apply to some datasets.
For instance Kriging generally requires as input a starting estimate of the error in the data. When this estimate is good the algorithm will perform well, but when the estimate is bad Kriging will often fail to converge. This should be considered a feature of the algorithm rather than a drawback. It is valuable that this method is capable of failure when its inputs do not conform to its assumptions because this means that it is more likely to raise a red flag in cases where performing a blind fit is dangerous. This is likely preferable to the silent failure of other algorithms, and as a result Kriging is the safest choice for exploring an unknown dataset.
In our benchmarking unfortunately this means that Kriging sometimes fails to provide an answer in a data-dependent fashion, which required providing it with somewhat more information about the errors than the other methods received, and moreover has much more variable performance as a function of the dataset parameters. However, Kriging has been used with considerable success for astronomical problems with errors sufficiently well understood to allow proper tuning \citep{Dumusque2017,ForemanMackey2017}.  For these implementations, the tuning process is typically strongly required and can be time-consuming, but the additional effort to choose an implementation suited to the individual regression problem being solved also produces better results.

\subsection{\algoname}
\label{subsec:ours}

In addition to existing algorithms, in this work we introduce \algoname~(\ac), a new regression algorithm designed to take full advantage of improvements in processing capability to produce improved bias and variance, particularly for large datasets such as those now produced in astronomical surveys.  When examined sufficiently locally, any differentiable function is approximately linear.  Therefore, we produce a predictor composed piecewise of linear approximations to the underlying function, using best linear unbiased estimators (BLUEs) and pruning the fit in a similar manner to MARS by removing breakpoints that are not statistically significant.  This algorithm assumes a set of relative errors $E_c$, and the python implementation includes a routine to estimate relative errors from the data if they are not user-provided.

\begin{algorithm}
\ac                    
\begin{algorithmic}
  \REQUIRE $N$ ordered triplets $(x_i,y_i,\xi_i)$, errors $E_c$. 
  \STATE Let ${H}\leftarrow \{\infty,\{\}\}$ store current score, predictor.
  \FOR{$e_c \in E_c$}
    \STATE Select at random $N/5$ ordered pairs $(x'_i, y'_i)$.
    \STATE Sort, such that $\forall\; i: x'_i \leq x'_{i+1}$.
    \STATE Let ${B} \leftarrow \{\{x_i, i=0 \ldots N/5-1\}\}$ be hypotheses
    \STATE Let ${Q} \leftarrow \{\}$ be the accepted hypothesis set.
    \WHILE{$\exists b\in B$}
      \STATE Remove $b$ from $B$.
      \STATE Let $\xi_i' \leftarrow e_c \xi_i$.
      \STATE Fit BLUEs to $b$, compute significance $\chi^2$ using $\{\xi_i'\}$.
      \IF{$\chi^2 < \mathrm{Cutoff}$}
        \STATE Let ${W} \leftarrow -1$ be the new break point.
        \FOR{$x_j \in b$}
          \IF{$|{x_{i<j}\in b}| > 4$ and $|{x_{i>j}\in b}| > 4$}
            \STATE Fit BLUEs to ${x_{i<j}\in b}$, compute $\chi^2_{-}$.
            \STATE Fit BLUEs to ${x_{i>j}\in b}$, compute $\chi^2_{+}$.
            \IF{$\chi^2_{-} + \chi^2_{+}  < \chi^2_W$ or $W == -1$}
              \STATE ${W} \leftarrow j$
            \ENDIF
          \ENDIF
        \ENDFOR
        \STATE Insert ${x_{i<W}\in b}$ and ${x_{i>W}\in b}$ into $B$.       
      \ELSE
        \STATE Place $b$ in $Q$.
      \ENDIF
    \ENDWHILE
    \STATE Let $H^\prime \leftarrow$ matched, piecewise BLUEs to $Q$.
    \STATE Let $R \leftarrow \sum_i{((y_i-H^\prime(x_i))/e_c \xi_i)^2}$.
	\IF{$R < H[0]$}
	  \STATE Let $H \leftarrow \{R, Q\}$.
	\ENDIF
  \ENDFOR
  \ENSURE Return $H[1]$.
\end{algorithmic}
\end{algorithm}

\ac~begins with the hypothesis that the data points $y_i$ are linearly related to $x_i$ with random noise superimposed linearly, with the goal of avoiding overfitting through only accepting complexity where the data can prove that it is required.  The default assumption is that the noise is normally distributed with zero mean and standard deviation $\xi_i$. A best linear unbiased least squares fit is then performed in order to evaluate the resulting $\chi^2$. If this $\chi^2$ allows \ac~to reject the assumption that the data is linear (given a cutoff) in favor of a piecewise linear function with a break then it will. In that case \ac~finds the datapoint such that independently fitting the data linearly on either side of this point produces the minimum total $\chi^2$ value. It then checks the $\chi^2$-value associated with each linear fit, and recursively searches until all fits fail to reject the linearity hypothesis.

If uncertainties on the data are not user-provided, they are estimated by performing local linear fits to windows of width $\sqrt{N}$ and evaluating the root-mean-squared residual. The result is interpreted as the standard deviation of a normal distribution. This biases the error estimates upward in the case where the function is not well resolved, but in the modelling limit this bias disappears. The cutoff $\chi^2$-value can be determined via cross-validation, though for numerical reasons it is better instead fix the cutoff and rescale the errors with a scale factor $e_c$, which is determined by cross-validation. For each scale factor this routine performs a fit on a random set of $N/5$ points with $\xi_i \rightarrow e_c \xi_i$. $\chi^2$ is then evaluated on the full dataset, selecting the fit which minimizes this across all $e_c$.

The comparison in individual fits is phrased in terms of $\chi^2$-values to take advantage of the fact that changing the cutoff $\chi^2$ value is equivalent to rescaling the uncertainties of the data. As a result, by fixing the cutoff at a convenient value and using cross-validation to rescale the errors appropriately it is only necessary to compare the observed $\chi^2$ to the median of the $\chi^2_{N-2}$ distribution
\begin{equation}
	M = k \left(1-\frac{2}{9k}\right)^3,
\end{equation}
where $k \equiv N-2$ is the number of degrees of freedom.

\ac~is fundamentally an algorithm for approximating the maximum likelihood solution in the space of piecewise continuous functions.
The effect of the cutoff on $\chi^2$ is to weight against more complex models in the prior likelihood.
Unfortunately we do not have a mathematical analysis of this algorithm demonstrating lack of bias or minimum variance\footnote{Indeed if~\ac misidentifies the number of breakpoints needed then it will likely be biased, though this statement is somewhat dependent on what is assumed about the underlying relationship.}.
What makes this difficult is largely that it selects its functions from a family parametrized by both continuous and discrete variables (e.g. the number of breakpoints) and so an analysis of its properties must take into account all possible conclusions it could draw about the number of parameters needed to express the underlying relationship.
Nevertheless, as we show in section~\ref{sec:results} this algorithm performs very well in practice and often outperforms other comparably-sophisticated methods.
Additionally, because any differentiable function is linear sufficiently locally and because we are producing a best linear unbiased estimator at every stage, it is reasonable to expect that the result of~\ac~will locally be a good fit, especially in the modeling regime where the data are dense.

A notable problem, however, is that \ac~will often produce a fit with a very large number of pieces when the underlying function is nonlinear and well-measured.  For example, a best-fit quadratic approximation to $y=(x-2)^2 - 3$ will have only three parameters, but \ac~will produce a piecewise many breakpoints, with the number of breakpoints increasing as there is more data.  However, we note that the key measure for interpolated regression is not a variance {\em per free parameter}, as it would be in determining the statistical significance of a fit.  Rather, the goal is simply to use the information available to produce the best possible predictor of the underlying function.  The true penalty in using \ac~to fit a quadratic is that the complex model with many breakpoints is difficult to interpret as quadratic, whereas fitting a quadratic will produce a result that is much more easily interpreted.

The implementations of \ac~used in our benchmarking trials are made available along with our testing suite and the other functions used.  The authors grant full license to use \ac~for any educational or non-profit purpose. 

\section{Nonparametric Regression for Two Errors}
\label{sec:2d}

As shown earlier, a truly two-error regression cannot be performed using one-error techniques.  The problem is that $E[X|Y]$ and $E[Y|X]$ will disagree with each other and do not characterize the true relationship between $X$ and $Y$.  This is because the underlying function has been convolved with a two-error error function, not merely a one-error one.  Thus, the goal is to decompose the observed probability distribution into an underlying function and such an error function.  With current techniques, this first requires solving the better-explored, yet no easier, problem of producing a parameterized probability density function in two dimensions given observations.  Once this is done, higher-level analysis may be applied to determine optimal fits from parameterized families of functions. 

\subsection{Gaussian Deconvolution}

Several off-the-shelf tools exist for parameterizing multi-dimensional probability distributions.  Early work focused on so-called kernel density estimation, which amounts to convolution of the data with a kernel representing the observational uncertainties but not the intrinsic scatter in the data\citep{rosenblatt1956}.  These methods also typically support estimating the total scatter and uncertainty by means of searching for a kernel which optimizes an objective function such as cross-validation likelihood\citep{Hall1992}.

More recently, Gaussian Deconvolution has emerged as a powerful technique for parameterizing multivariate density functions which have Gaussian-like tails\citep{bovy2011}. This technique, which has an off-the-shelf implementation available at~\url{https://github.com/jobovy/extreme-deconvolution}, takes as input a desired number of Gaussians and a dataset with known covariance matrices corresponding to observational errors, rather than intrinsic scatter.  It produces as output the parameters of the requested number of Gaussians, such that those Gaussians when convolved with normally distributed errors maximize the likelihood of the data.

There are two key advantages of Gaussian Deconvolution over ordinary kernel density estimation. First, it robustly accounts for known observational uncertainties, and so avoids overestimating the intrinsic scatter.  Second, for large datasets it produces parametrization with a controlled number of degrees of freedom, whereas kernel density estimation produces a distribution with $N$ degrees of freedom given $N$ data points. As a result, Gaussian Deconvolution generally produces output in a good format for further processing and statistical analysis, whereas it can be far more difficult to use the output from kernel density estimation.  However, it does fundamentally assume normally-distributed errors, which are most common in astronomy but by no means the only ones found in nature.

One additional difficulty associated with Gaussian Deconvolution is that it does not a priori have a preference over the number of Gaussians it produces: this is a user specified parameter.  In this work we have addressed this by means of a distance heuristic.
We begin by evaluating the deconvolution for a single Gaussian.  We then increase the number of Gaussians used by one, and evaluate the statistical distance between the previous fit and the new fit, as measured by the Bhattacharyya distance\citep{bhtc}
\begin{equation}
	D(p,q) = -\ln \int \sqrt{p(\boldsymbol{x}) q(\boldsymbol{x})}d^2 x.
\end{equation}
If this distance exceeds a threshold, which we set to $0.1$, the distribution is still changing significantly as a function of the number of Gaussians and so we continue increasing this number, as the parametrization is likely underfit.
We halt the process when this criterion is no longer satisfied.
Our choice of distance metric and cutoff is somewhat arbitrary.
The metric ought to be something which vanishes when the distributions are identical and increases as they differ more strongly in regions of significant support.
The cutoff controls a preference between bias and variance.
More sophisticated choices are possible by cross-validation and likelihood optimization but these tend to have unacceptably long runtimes, and in practice we found our choices to perform well.

\subsection{Producing a 2-E Predictor}

\begin{figure}[h]
	\plotone{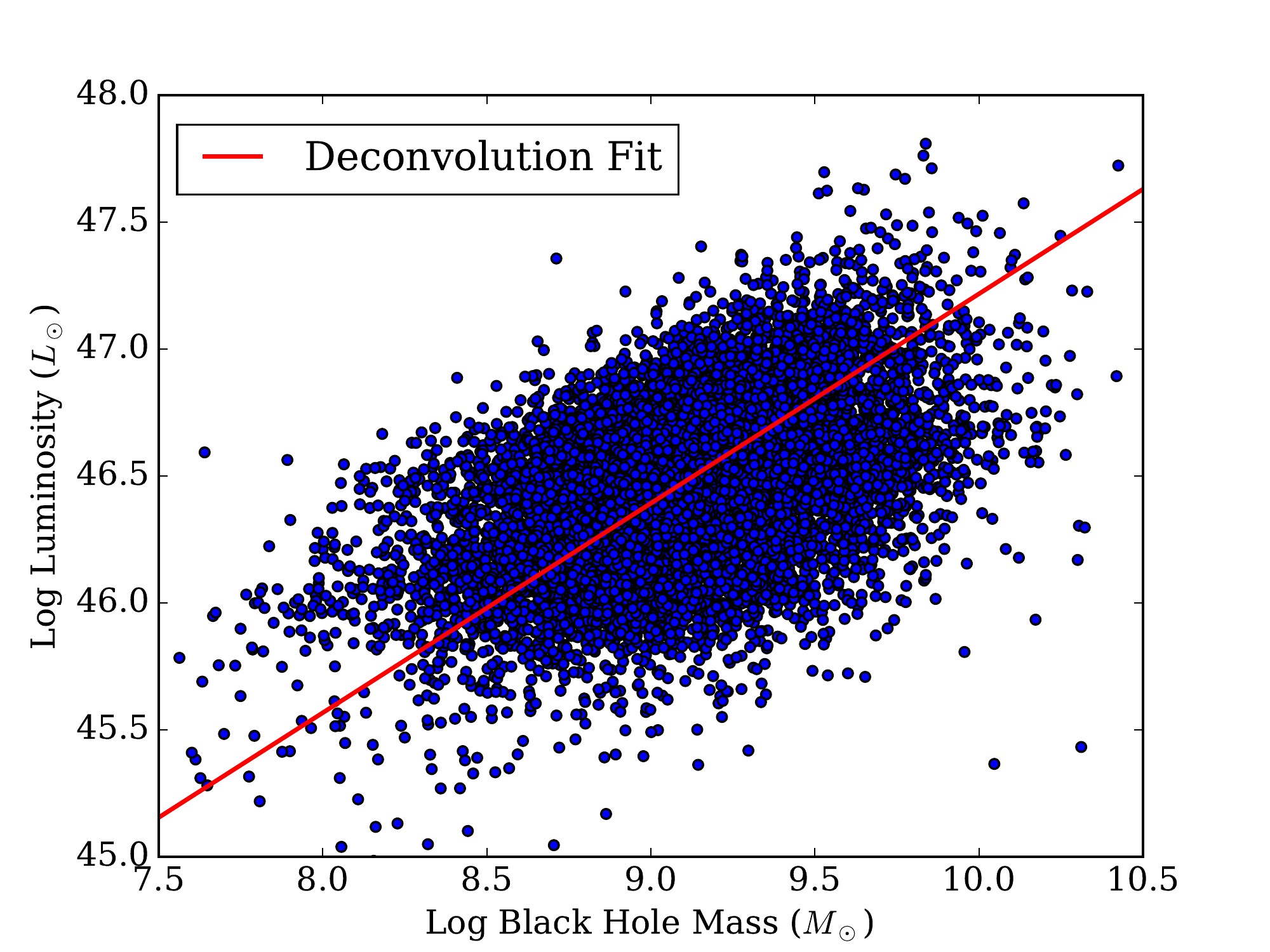}
	\caption{Best-fit linear predictor for the quasar mass-luminosity plane first shown in Fig. \ref{fig:qsomeds}.  A Gaussian deconvolution finds that the distribution is well-fit by one Gaussian, and the best-fit predictor is therefore a line with $L \propto M^{0.8}$ (red).  Note that the one-error fits with $L \propto M^{1.7}$ and $L \propto M^{0.4}$ were both poor. An arbitrary 5\% of the data was used in the fit for performance purposes.}
	\label{fig:qsogauss}
\end{figure}

In some cases, the Gaussian deconvolution described above will produce a good fit with just one Gaussian.  In that case, the best-fit predictor will be a linear relationship using one of the two principal axes of that Gaussian, with the other noise.  For example, this produces a good fit to the quasar distribution discussed earlier (Fig. \ref{fig:qsogauss}), even though the one-error fits with $L \propto M^{1.7}$ and $L \propto M^{0.4}$ were both poor.

\begin{figure}[h]
	\plotone{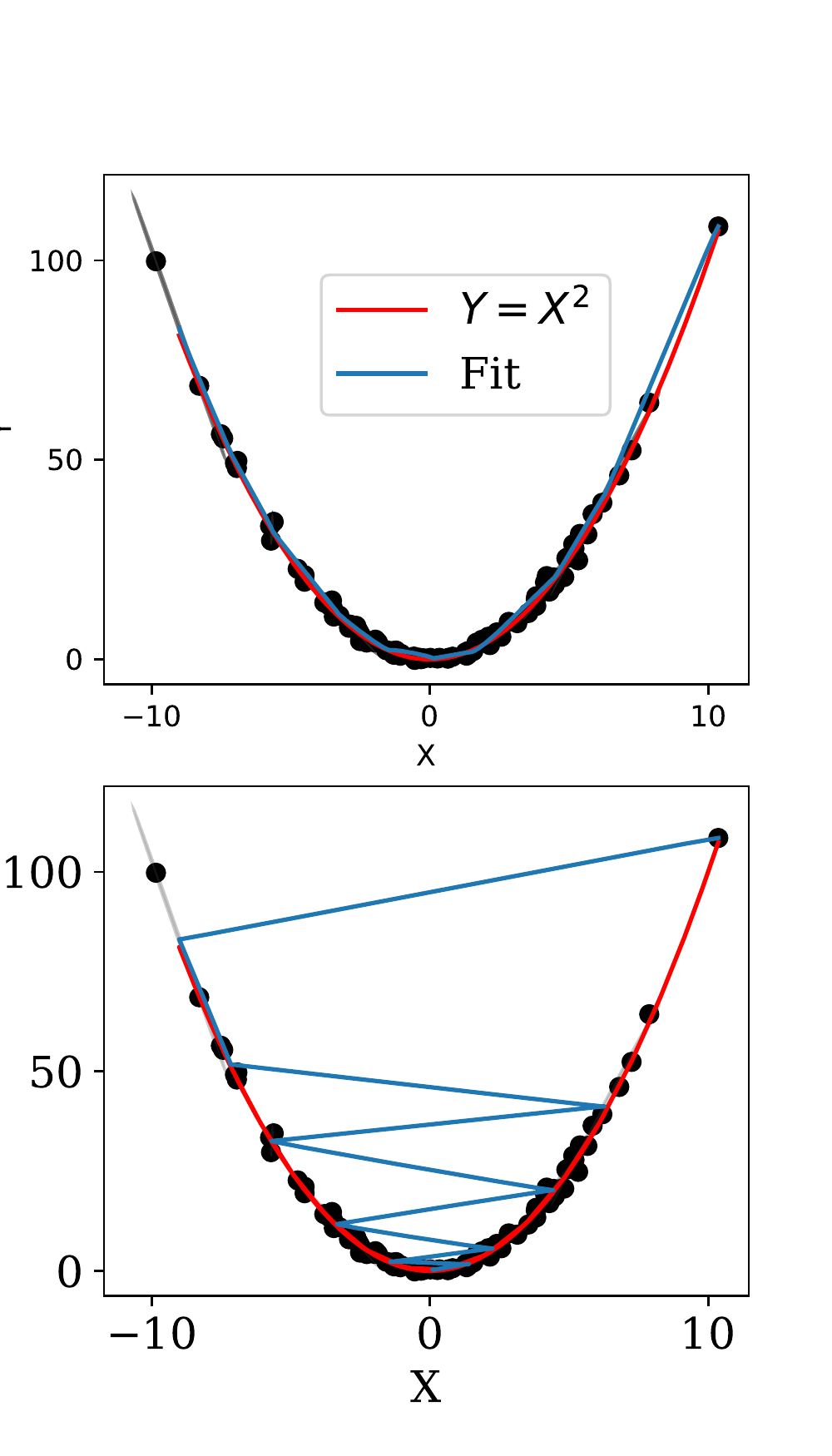}
	\caption{(Top) A cubic spline (blue) connecting Gaussian deconvolution components (grey) calculated from a set of points (black) drawn from an underlying parabolic function, $Y = X^2$ (red), with the components connected in the correct order, monotonic in $X$.  (Bottom) The same spline with the component connected in an incorrect order, monotonic in $Y$. The Gaussian components are quite small, of comparable width to the data points, but one is clearly visible on the top-left. The former is a very good fit to the original parabola. The latter is a good fit over half of the parabola but fails to connect the other half of the curve.}
	\label{fig:parabola}
\end{figure}

However, for more complicated cases, deconvolution will produce multiple Gaussians which must then be combined into a predictor that can be used for interpolation.  Intuitively, a predictor should go through the central points of these Gaussians and should locally point along one of the principal axes.  Thus, a natural candidate is a cubic spline, with each Gaussian component fixing a point and derivative, and the remaining parameter chosen to smoothly connect them.  For well-behaved data, this can be an effective predictor for the underlying function (Fig. \ref{fig:parabola}a).
However, there is an inherent ordering problem that must be solved: if the Gaussian components are connected in the incorrect order, a very poor predictor is instead produced (Fig. \ref{fig:parabola}b).

\subsection{\algonametwod}

The ordering problem can be very difficult for some underlying functions.  A monotonic underlying function can often be approximated by connecting the Gaussian component monotonically in either variable.  However, fitting points drawn from the parabola $Y = X^2$ will produce a reasonable result when components are connected monotonically in $X$, but a very poor one when they are connected monotonically in $Y$ (Fig. \ref{fig:parabola}).  An ellipse would poorly fit if connected monotonically in either variable.  

The best solution involves using additional knowledge about the underlying function to pick an ordering. However, in the absence of any such knowledge, a good generalized default answer is to find a minimal length/maximum likelihood connected path containing the centers of each Gaussian component, then to use that as an ordering. Once more we use the Bhattacharyya distance, this time as a proxy for the path likelihood between two Gaussian components. In practice, for speed of execution we then approximate the maximum likelihood path from a minimal spanning tree in our implementation of a 2-E nonparametric regression, \algonametwod (\actwod):

\begin{algorithm}
	\actwod\                     
	\begin{algorithmic}                    
		\REQUIRE $N$ ordered pairs $(X_i,Y_i)$, errors $(\Delta~X_i,\Delta~Y_i)$cutoff distance $\delta_c$, isVertical (boolean)
		\STATE $\delta \leftarrow 2 \delta_c$
		\STATE $N \leftarrow 1$
		\STATE $G^\prime \leftarrow $ = Gaussian Deconvolution ($N$ components)
		\WHILE{$\delta > p_c$}
			\STATE $G \leftarrow G^\prime$
			\STATE $N \leftarrow N + 1$
			\STATE $G^\prime \leftarrow$ = Gaussian Deconvolution ($N$ components)
			\STATE $\delta \leftarrow $ Distance$(G,G^\prime)$
		\ENDWHILE
		\STATE For each component $G_i$, let the mean lie at $(x_i,y_i)$		
		\FOR{$i = 0 \ldots N-1$}
			\FOR{$j = 0 \ldots N-1$} 
				\STATE $D[i,j] \leftarrow$ Distance$(G_i,G_{i'}^\prime)$
			\ENDFOR
		\ENDFOR
		\IF{isVertical}
			\STATE Start $\leftarrow j|\textrm{min}(y_j)$
			\STATE Stop $\leftarrow j|\textrm{max}(y_j)$
		\ELSE
		\STATE Start $\leftarrow j|\textrm{min}(x_j)$
		\STATE Stop $\leftarrow j|\textrm{max}(x_j)$
		\ENDIF
		\STATE $DG \leftarrow$ Graph with adjacency matrix $D$
		\STATE $T \leftarrow$ Minimum Spanning Tree($G$)
		\STATE $P \leftarrow$ Shortest Path (Start $\rightarrow$ Stop) through T
		\STATE $P_{XY} \leftarrow$ Hermite Spline through $P$, with derivatives given by each $G_i$ passed through
		\ENSURE Return $P_{XY}$                
	\end{algorithmic}
\end{algorithm}

It should be noted that for particularly difficult underlying functions to fit, the minimal spanning tree may require that a node is visited multiple times on the path $P$.  For the purposes of the spline, we only consider points in the order that they are first visited, passing through each once.  However, it is important to remember that unlike what is possible for 1-E regressions, \actwod\ \ does not guarantee a minimum bias/variance fit.  If there is additional information about the likely underlying function, using this to force a specific starting point, stopping point, or full path is likely to produce a better predictor than the fully nonparametric version implemented in \actwod.  

Moreover, the entire basis for a predictor is an assumption that the underlying function is sufficiently smooth when compared with both measurement errors and the sampling density that it is possible to use nearby points to predict an unmeasured value.  If the best path for a given dataset is often self-intersecting, it is likely an indication that this is not true for that particular problem and that the predictor carries little meaning.  Because \actwod\ \ and several of the tools described for one-error regressions will always produce a result, even in cases where that result is not meaningful, astronomers using these tools must be careful to exercise their own judgment in determining whether the underlying relation produced is valid and whether any physical meaning should be ascribed to it.

\section{Benchmarking Tests}
\label{sec:tests}

The benchmarks were run on a workstation with an Intel Core i7 processor.  All tests were performed in RAM with no other significant processes running.  The tests were implemented in Python 3.5 with NumPy 1.11.1 and SciPy 0.17.1. Medians and other binned regressions were implemented in vectorized form, maximizing calls to underlying Fortran and C components of NumPy. Boosted Tree regression was implemented with the Scikit-learn 0.17.1 Python module. The Multivariate Adaptive Regression Spline technique was implemented with the Py-Earth 0.1.0 Python module. \ac\ was implemented in with SciPy spline fitting routines. 

We apply each of these techniques to the following generalized procedure:
\begin{enumerate}
\item{Pick a underlying model function $Y(X)$. We used a linear function, a linear superposition of sinusoids, and a square wave with several periods.}
\item{Over $k=100$ trials:
    \begin{itemize}
    \item{Choose $N$ points from a uniform random distribution over the domain $x\in \left[0,1\right]$ and evaluate the underlying model function at those points.}
    \item{For each chosen point, draw a noise correction from a Gaussian error distribution with variance given by the mean and add it to the function value.}
    \item{Run the regression technique, recording the result and running time}
    \end{itemize}}
\item{Determine the average bias, variance, and running time for each technique.}
\end{enumerate}
This procedure was repeated for $N=100$ and $N=10^5$ to simulate a smoothing problem and a modeling problem respectively.
These numbers were chosen according to Eq.\ \eqref{eq:decision} given the sinusoidal function.
We also compared different bin widths for binned medians with $N=10^5$.
In all cases points were chosen uniformly over the range, and the bias, variance, etc. were evaluated on the inner $50\%$ of the range.

\section{Benchmarking Results}
\label{sec:results}

\subsection{Smoothing}

Table\ \ref{tab:bench1} shows the results of our benchmarks with $N=100$.
This corresponds to a smoothing problem.  For the sinusoid, connecting the dots, means, and medians do best on bias by a wide margin, at the cost of somewhat worse performance on variance. Their good performance on bias is to be expected, as they are unbiased. Connect the dots performs particularly poorly on variance because it uses purely local information, and in that sense is neither a fitting nor smoothing method at all.
Cubic splines also performs poorly on variance, likely because without guidance on the scatter it is susceptible to overfitting.
Of the modeling algorithms applied to this smoothing problem, Random Forest performs best on bias while Boosted Trees performs best on variance. The difference between these two likely reflects a different choice of bias-variance tradeoff rather than a fundamental difference.

Linear smoothing presents the simplest problem, and biases are quite low across the board.
This is because the interpolation error vanishes, so the bias that remains is a combination of random noise associated with doing a finite number of trials and intrinsic bias in the method.
Disentangling the two is difficult, as the more sophisticated methods may have complicated behavior in terms of how rapidly the random portion of the bias measurement averages out.
This can be seen in the performance of binned medians and means, which achieve significantly lower bias than the other methods despite several of the other methods being unbiased.

On variance Kriging and MARS performed best, followed by Local Linear regression.  This is as might be expected, as all three methods are in one fashion or another looking to fit as few basis functions to the data, and this data is fundamentally linear. Similar behavior is seen for the square wave, though here \ac~and Boosted Trees have an easier time than the other modeling methods.

Overall, the smoothing algorithms (running medians and means and binned medians) perform quite well on these smoothing problems.  They also hold a considerable advantage in runtime, and in many cases are limited by Python function call overhead rather than the underlying algorithm, though none of the algorithms were unacceptably slow.  For example, our simple implementation of binned medians had longer runtime than running medians due to our choice of specific Python libraries called.

There is not a clear advantage to using modeling algorithms, except perhaps for Boosted Trees and MARS, which leverage higher bias and runtime in exchange for a lower variance and hence a more robust answer.
In cases where low RMSE is the main goal the two local regression techniques are also quite good.

It should be noted that the biggest improvement between the sinusoid and the linear function was in the modeling algorithms.
This is because a linear function is always a modeling problem for these methods, as all of them have linear functions in their basis sets.

\subsection{Modeling}

Table\ \ref{tab:bench2} shows the results of our benchmarks with $N=10^5$.
This point density makes this a modeling problem in both cases.
For performance reasons Kriging was omitted from these runs, as the runtime of the implementation we used scales much more than linearly with the number of points.

For the linear function, the primary difference is that the modeling algorithms do a much better job of incorporating the additional information that comes with greater point density. In particular, MARS and~\ac~achieve significantly lower variance than the other methods. They can determine that all of the points should be used together in producing one linear fit. In almost every case both the bias and variance of the modeling algorithms fell. 

Overall, MARS performed best on bias and RMSE and tied with \ac~for best on variance. \ac~ also achieved the second best RMSE. Some of the smoothing algorithms improved in bias, primarily as a result of increased window and bin sizes. However, their variance was already limited by the local scatter, and thus did not improve.

For the more complex sinusoid, the advantage of modeling is more apparent.  Although some of the smoothing algorithms achieve low bias, none of them achieves variance comparable to the better modeling algorithms like Local Quadratic regression and MARS.  This suggests that the modeling algorithms are actually uncovering the underlying function, as their name suggests, whereas the smoothing algorithms are just producing something with a bounded derivative regardless of if it models the function well. Of the modeling algorithms, Random Forest, \ac~and MARS performed best on bias on this test, and MARS was the clear winner on variance.

The square wave is an interesting case, as here \ac~performs worse than expected.
This is because for fair comparison we used scatter-based error estimates rather than providing observational uncertainties.
The method \ac~uses to estimate these errors systematically overestimates in the vicinity of discontinuities, resulting in a fit which smooths over these regions.
Thus while it achieves the best bias of any other modeling algorithm other than Random Forest, it is in the middle of the pack for variance.

In all cases runtimes increased with increasing $N$, but Kriging, \ac~and MARS scale more poorly than the other algorithms. In the case of Kriging this is because the covariance matrix increases in complexity rapidly with $N$.  For \ac~it is because it considers all datapoints to be potential breakpoints, whereas the other algorithms either do not couple choice of breakpoint to the observation locations or apply simple smoothing procedures. For MARS, the increase in runtime is only evident for the sinusoid. This is because MARS searches for simple descriptions of the data, and so for simple data it quickly finds a simple description whereas for more complicated functions it has to incorporate more complexity and will perform more similarly to \ac.

\subsection{Binning}

Finally, Table\ \ref{tab:bench3} shows the results of these trials using binned medians for various bin widths.  The bias and variance depend upon the bin width selected and how it compares with the underlying function.  For the linear case, broader bins will perform better, as the underlying linear approximation holds and more points are used in the measurements.  For the more complex sinusoid, aliasing of the bin edges with the phase of the sinusoid can cause individual bins to be strongly biased.  

The difference between the two illustrates the difficulty in picking a single bin size for all problems, and of course there is the additional difficulty that a uniform bin size is not generally applicable even to a single function.  Still, for the correct set of bins, the underlying function is reasonably well approximated.  In other words, it is not the concept of binned medians that is flawed, but rather the ability of astronomers to guess a priori how their data should be optimally binned when faced with a new regression problem, and the impossibility of picking one-size-fits-all bins even for a single problem.  The idea of modeling algorithms is that they use the data itself to decide which data points matter for the function value at each point, effectively producing an intelligent choice of bins, sometimes of variable width, that closely approximate the curvature of the underlying function.  

These are but a handful of a large set of possible tests, designed to illustrate two key principles.  First, there is a real difference between smoothing and modeling problems, and the right algorithm will provide a meaningful improvement.
Second, there is a tradeoff between statistical performance on modeling problems and runtime, and with runtimes pushed low by advances in computing there is increased capability to employ sophisticated methods.

\begin{table}
\caption{Table of bias, variance, root-mean-squared error, and runtime for seven different fitting algorithms applied to two different underlying functions mixed with Gaussian noise scaled with Poisson variance. $N=100$ points were used over the range $[0,1.0]$ with 100 trials. Metrics are evaluated by numeric integration on the inner 50\% of the range.}
\label{tab:bench1}
\begin{tabular}{lrrrr}
\hline
                     &        $B$ &   $\sigma^{2}$ &       RMSE &   Time (s) \\
\hline
\hline\\\multicolumn{5}{c}{$y=4 + \sin(2\pi x) + \sin(4 \pi (x+0.1)) + \cos(6\pi x)+ \cos(14\pi x)$}\\\hline\\
 Connect the Dots    &     0.0010 &       0.6618 &     0.8135 &     0.0001 \\
 Means               &     0.0184 &       0.1752 &     0.4189 &     0.0003 \\
 Medians             &     0.0412 &       0.2105 &     0.4607 &     0.0003 \\
 Binned Medians (10) &     0.1357 &       0.2812 &     0.5474 &     0.0007 \\
 Boosted Trees       &     0.0812 &       0.1471 &     0.3921 &     0.0111 \\
 Random Forest       &     0.0011 &       0.5520 &     0.7430 &     0.0103 \\
 Local Linear        &     0.1127 &       0.1530 &     0.4070 &     0.0006 \\
 Cubic Spline        &     0.0021 &       4.0878 &     2.0218 &     0.0008 \\
 MARS                &     0.1740 &       0.1773 &     0.4557 &     0.0044 \\
 Local Quadratic     &     0.0624 &       0.1811 &     0.4301 &     0.0134 \\
 Kriging             &     0.0372 &       0.2629 &     0.5140 &     0.3262 \\
 ZeBRA               &     0.0973 &       0.1963 &     0.4536 &     0.3691 \\
\hline\\\multicolumn{5}{c}{$y=x$}\\\hline\\
 Connect the Dots    &     0.0161 &       0.6331 &     0.7958 &     0.0001 \\
 Means               &     0.0149 &       0.0937 &     0.3065 &     0.0003 \\
 Medians             &     0.0103 &       0.1370 &     0.3703 &     0.0003 \\
 Binned Medians (10) &     0.0142 &       0.1036 &     0.3222 &     0.0007 \\
 Boosted Trees       &     0.0187 &       0.0484 &     0.2209 &     0.0124 \\
 Random Forest       &     0.0213 &       0.5338 &     0.7309 &     0.0108 \\
 Local Linear        &     0.0031 &       0.0283 &     0.1683 &     0.0007 \\
 Cubic Spline        &     0.0047 &       0.0328 &     0.1813 &     0.0001 \\
 MARS                &     0.0040 &       0.0173 &     0.1316 &     0.0024 \\
 Local Quadratic     &     0.0060 &       0.0470 &     0.2169 &     0.0177 \\
 Kriging             &     0.0028 &       0.0162 &     0.1272 &     0.2881 \\
 ZeBRA               &     0.0060 &       0.0922 &     0.3037 &     0.3536 \\
\hline\\\multicolumn{5}{c}{$y=\mathrm{sgn}(\sin(10\pi x))$}\\\hline\\
 Connect the Dots    &     0.0357 &       0.7236 &     0.8514 &     0.0001 \\
 Means               &     0.0401 &       0.2323 &     0.4836 &     0.0003 \\
 Medians             &     0.0024 &       0.2598 &     0.5097 &     0.0004 \\
 Binned Medians (10) &     0.0479 &       0.2028 &     0.4529 &     0.0010 \\
 Boosted Trees       &     0.0261 &       0.2585 &     0.5091 &     0.0160 \\
 Random Forest       &     0.0043 &       0.6287 &     0.7929 &     0.0111 \\
 Local Linear        &     0.0930 &       0.2928 &     0.5490 &     0.0007 \\
 Cubic Spline        &     0.0055 &       3.6546 &     1.9117 &     0.0008 \\
 MARS                &     0.0589 &       0.2788 &     0.5313 &     0.0036 \\
 Local Quadratic     &     0.0436 &       0.3130 &     0.5612 &     0.0156 \\
 Kriging             &     0.0048 &       0.3328 &     0.5769 &     0.2750 \\
 ZeBRA               &     0.0169 &       0.2802 &     0.5296 &     0.3301 \\
\hline
\end{tabular}
\end{table}

\begin{table}
\caption{Table of bias, variance, root-mean-squared error, and runtime for seven different fitting algorithms applied to two different underlying functions mixed with Gaussian noise scaled with Poisson variance. $N=10000$ points were used over the range $[0,1.0]$ with 100 trials. Metrics are evaluated by numeric integration on the inner 50\% of the range.}
\label{tab:bench2}
\begin{tabular}{lrrrr}
\hline
                     &        $B$ &   $\sigma^2$ &       RMSE &   Time (s) \\
\hline
\hline\\\multicolumn{5}{c}{$y=4 + \sin(2\pi x) + \sin(4 \pi (x+0.1)) + \cos(6\pi x)+ \cos(14\pi x)$}\\\hline\\
 Connect the Dots    &     0.0001 &       0.6674 &     0.8169 &     0.0006 \\
 Means               &     0.0004 &       0.0985 &     0.3138 &     0.0024 \\
 Medians             &     0.0025 &       0.1316 &     0.3628 &     0.0045 \\
 Binned Medians (10) &     0.1365 &       0.1631 &     0.4263 &     0.0021 \\
 Boosted Trees       &     0.0713 &       0.0906 &     0.3094 &     0.0587 \\
 Random Forest       &     0.0015 &       0.5106 &     0.7146 &     0.1796 \\
 Local Linear        &     0.0512 &       0.0879 &     0.3009 &     1.9729 \\
 Cubic Spline        &     0.0369 &      17.1262 &     4.1385 &     2.9844 \\
MARS                &     0.0003 &       0.0074 &     0.0860 &     7.9865 \\
 Local Quadratic     &     0.0178 &       0.0442 &     0.2111 &    18.6860 \\
 ZeBRA               &     0.0021 &       0.0977 &     0.3126 &   323.7618 \\
\hline\\\multicolumn{5}{c}{$y=x$}\\\hline\\
 Connect the Dots    &     0.0070 &       0.6694 &     0.8182 &     0.0005 \\
 Means               &     0.0004 &       0.0965 &     0.3107 &     0.0023 \\
 Medians             &     0.0007 &       0.1322 &     0.3636 &     0.0044 \\
 Binned Medians (10) &     0.0010 &       0.0011 &     0.0337 &     0.0020 \\
 Boosted Trees       &     0.0008 &       0.0019 &     0.0435 &     0.0573 \\
 Random Forest       &     0.0060 &       0.5163 &     0.7186 &     0.1503 \\
 Local Linear        &     0.0004 &       0.0007 &     0.0260 &     1.7038 \\
 Cubic Spline        &     0.0005 &       0.0003 &     0.0175 &     0.0025 \\
 MARS                &     0.0002 &       0.0002 &     0.0123 &     0.0961 \\
 Local Quadratic     &     0.0019 &       0.0011 &     0.0328 &    15.9194 \\
 ZeBRA               &     0.0009 &       0.0002 &     0.0145 &   259.9066 \\
\hline\\\multicolumn{5}{c}{$y=\mathrm{sgn}(\sin(10\pi x))$}\\\hline\\
 Connect the Dots    &     0.0033 &       0.6633 &     0.8145 &     0.0006 \\
 Means               &     0.0014 &       0.0974 &     0.3121 &     0.0021 \\
 Medians             &     0.0011 &       0.1319 &     0.3632 &     0.0040 \\
 Binned Medians (10) &     0.0318 &       0.0892 &     0.3004 &     0.0020 \\
 Boosted Trees       &     0.0586 &       0.2260 &     0.4790 &     0.0522 \\
 Random Forest       &     0.0020 &       0.5220 &     0.7225 &     0.1573 \\
 Local Linear        &     0.0471 &       0.1396 &     0.3766 &     1.6232 \\
 Cubic Spline        &     0.0170 &       4.8323 &     2.1983 &     3.0145 \\
 MARS                &     0.0062 &       0.0296 &     0.1720 &    28.8410 \\
 Local Quadratic     &     0.0260 &       0.0675 &     0.2610 &    14.9648 \\
 ZeBRA               &     0.0040 &       0.1418 &     0.3766 &   284.8095 \\
\hline
\end{tabular}
\end{table}

\begin{table}
\caption{Table of bias, variance, root-mean-squared error, and runtime for binned medians applied with different numbers of bins and two different underlying functions mixed with Gaussian noise scaled with Poisson variance. $N=10000$ points were used over the range $[0,1.0]$ with 100 trials. Metrics are evaluated by numeric integration on the inner 50\% of the range.}
\label{tab:bench3}
\begin{tabular}{lrrrr}
\hline
     &        $B$ &   $\sigma^2$ &     RMSE &   Time (s) \\
\hline
\hline\\\multicolumn{5}{c}{$y=4 + \sin(2\pi x) + \sin(4 \pi (x+0.1)) + \cos(6\pi x)+ \cos(14\pi x)$}\\\hline\\
 3   &     0.1150 &       0.1193 &     0.3641 &     0.0015 \\
 4   &     0.0469 &       0.1160 &     0.3437 &     0.0015 \\
 5   &     0.1726 &       0.1392 &     0.4111 &     0.0016 \\
 6   &     0.2317 &       0.1414 &     0.4417 &     0.0017 \\
 10  &     0.1372 &       0.1609 &     0.4239 &     0.0019 \\
 25  &     0.0234 &       0.0170 &     0.1324 &     0.0028 \\
 50  &     0.0061 &       0.0065 &     0.0812 &     0.0042 \\
 100 &     0.0007 &       0.0105 &     0.1025 &     0.0071 \\
\hline\\\multicolumn{5}{c}{$y=x$}\\\hline\\
 3   &     0.0240 &       0.0006 &     0.0344 &     0.0014 \\
 4   &     0.0125 &       0.0005 &     0.0259 &     0.0015 \\
 5   &     0.0064 &       0.0005 &     0.0241 &     0.0016 \\
 6   &     0.0026 &       0.0006 &     0.0251 &     0.0016 \\
 10  &     0.0010 &       0.0011 &     0.0330 &     0.0019 \\
 25  &     0.0014 &       0.0028 &     0.0525 &     0.0028 \\
 50  &     0.0009 &       0.0051 &     0.0715 &     0.0042 \\
 100 &     0.0009 &       0.0106 &     0.1030 &     0.0069 \\
\hline\\\multicolumn{5}{c}{$y=\mathrm{sgn}(\sin(10\pi x))$}\\\hline\\
 3   &     0.0004 &       0.2551 &     0.5051 &     0.0014 \\
 4   &     0.0047 &       0.2662 &     0.5160 &     0.0015 \\
 5   &     0.0004 &       0.2547 &     0.5046 &     0.0016 \\
 6   &     0.0011 &       0.2606 &     0.5105 &     0.0017 \\
 10  &     0.0300 &       0.0881 &     0.2984 &     0.0020 \\
 25  &     0.0036 &       0.0575 &     0.2397 &     0.0029 \\
 50  &     0.0084 &       0.0229 &     0.1517 &     0.0043 \\
 100 &     0.0021 &       0.0202 &     0.1422 &     0.0070 \\
\hline
\end{tabular}
\end{table}

\section{Discussion}
\label{sec:discussion}

We have considered several algorithms for nonparametric regression on bivariate data.  The most commonly used techniques in Astronomy are variations of binned or running medians and averages.  Other algorithms considered include Multivariate Adaptive Regressive Splines (MARS), Boosted Trees, Random Forests, Local Linear and Local Quadratic regression, Kriging, and \algoname~(\ac), a new method developed in this work.  Each of these methods was benchmarked on a series of regression problems with different properties in order to help determine which is optimal.

\subsection{Which One Should I Pick? (1-E)}

\begin{figure}[!ht]
\plotone{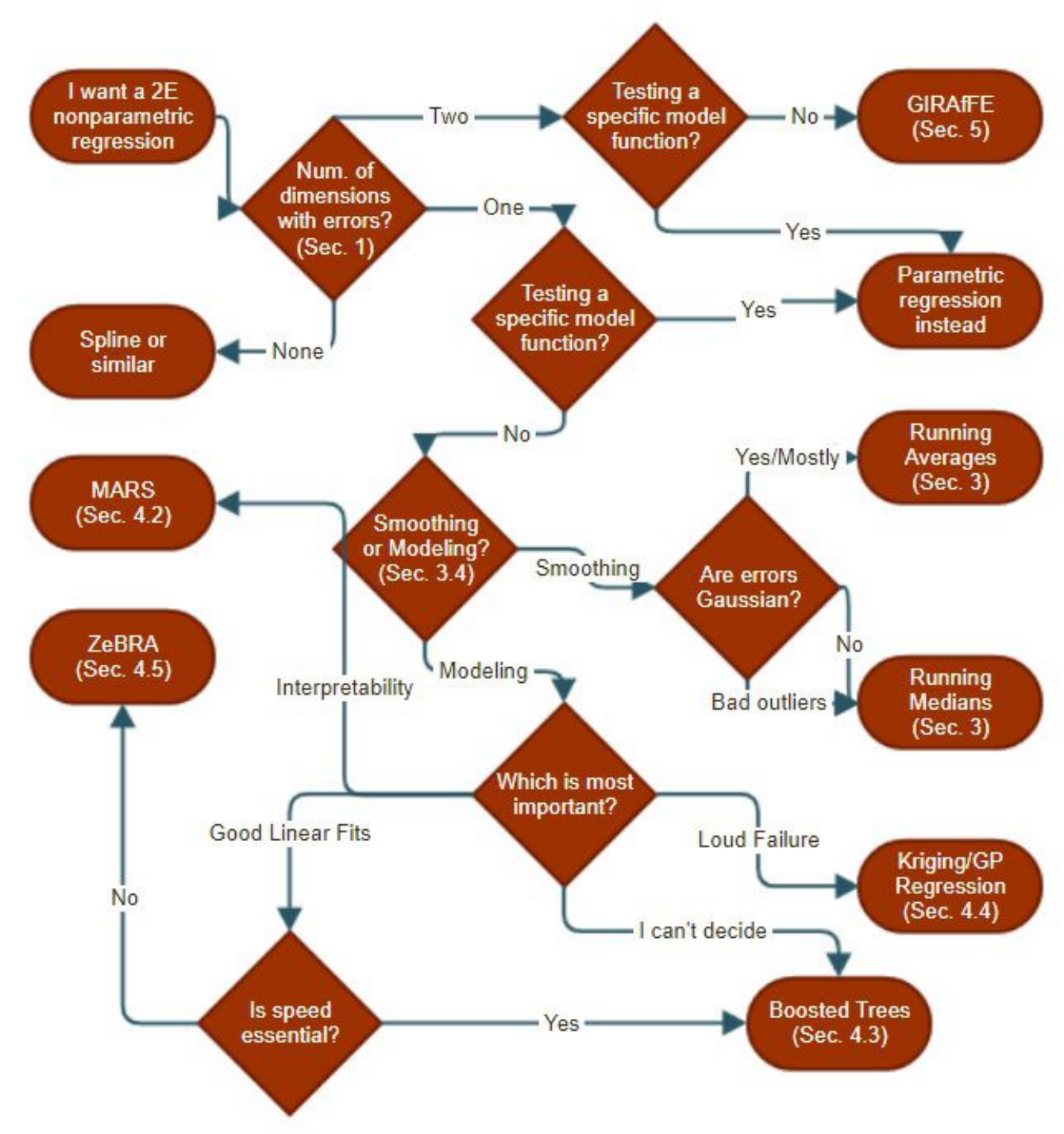}
\caption{Flowchart summarizing recommendations for how to choose the best nonparametric regression technique.  The correct choice depends upon the details of the particular problem being solved, so it is difficult to provide a default recommendation.}
\label{fig:flowchart}
\end{figure}

In choosing numerical methods there are always tradeoffs, and so we do not advocate for any particular method to the exclusion of all others.  We summarize our recommendations for choosing a technique in Fig. \ref{fig:flowchart}. 
For many common use cases, particularly where it is unclear what ought to be prioritized, the flowchart will terminate with Boosted Trees.  Other algorithms are better optimized for minimum bias (\ac), safety (Kriging), and interpretability (MARS) where one of those outweighs the others.

A particularly strong motivation for this work is the evolution of astronomy towards increasingly large datasets.  This means that astronomers should transition from smoothing towards modeling algorithms.  Depending upon the specific usage, it may help to consider several factors when deciding which method will perform best.  In some cases our recommendations are necessarily subjective based upon our experiences installing and using these techniques in our benchmarking process.  In summary, for large dataset modeling problems: 
\begin{itemize}
\item{Fit Quality: Benchmarking indicates that MARS, Boosted Trees, and \ac~produced the best fits, with the best choice dependent upon the underlying function.  \ac~is designed to perform best on functions with large linear components, MARS produced a much better fit than other algorithms on the composite sinusoid, and Boosted Trees was most successful at dealing with the discontinuities present in the square wave.  MARS had marginally the best overall fits, but often at the cost of a higher bias.  Kriging will produce a high quality fit when it converges, but will sometimes fail.  Medians and averages consistently produce the worst fits, and can be unacceptably poor even for simple functions.}
\item{Interpretability: When the underlying function is simple, MARS will produce a correspondingly simple predictor if the underlying function is part of its basis set.  Boosted Trees will similarly produce relatively few breakpoints.  At the other end of the scale, \ac~and running medians are entirely empirical and will always produce numerical approximations to even simple underlying function, with \ac~producing better approximations.}
\item{Non-Biased: \ac~is both designed to be unbiased and has the minimum bias in many of the practical applications tested.  Boosted Trees and MARS are not unbiased, but practically have low bias in typical cases.  Kriging will be unbiased when its underlying assumptions hold.  Running medians are formally biased for finite datasets, and averages biased for even infinitely large datasets.  However, in practice both yield reasonably unbiased results in the modeling regime, as these flaws become most apparent either when smoothing or in the very large-$N$ limit, as both take a large number of points to converge to their minimum theoretical bias.}
\item{Ease of Use: Running medians is certainly the simplest of these algorithms to implement from scratch.  Boosted Trees exists as part of the scipy package in python and worked immediately.  It is our hope in releasing an implementation of \ac\ that it will be similarly straightforward to use, although it is currently not part of any existing packages and thus requires a fresh installation.  Kriging also has a python implementation that worked quickly.  MARS has been implemented in python as py-earth, but has several complex dependencies.  The authors were ultimately unable to run MARS in python 3.x, and a python 2.7 version was only successfully installed on one of three computers on which it was attempted.  Once properly installed, each algorithm has a straightforward function call producing output in a reasonable format.}
\item{Safety: Ideally, an algorithm should produce not just a fit but also a diagnostic of the goodness of fit. Unfortunately, goodness of fit is algorithm-specific and may not be well-defined. For example, MARS increases complexity until a good fit is achieved, and so statistical goodness of fit is guaranteed while simplicity of fit is the real indicator of how well the model reflects the underlying relationship. Thus, in practice what is desirable is that an algorithm should fail noisily when the assumptions underlying its fit are found not to hold, and Kriging is the only algorithm considered that does so.  MARS, Boosted Trees, Random Forests and \ac\ can be considered to have failed when they produce a very large number of uncorrelated breakpoints, although the algorithms do not explicitly check for this. Running medians have no easily-checked failure condition that we have found.}
\item{Time: Runtime for all of these algorithms other than Kriging is short for small or moderate samples, Even \ac\ required only a few minutes for 10,000 points on our test machine, which would be acceptable for most analyses that do not need to be run repeatedly on different large samples or in real-time.  The simpler algorithms have runtimes that only depend upon the size of the sample, but more sophisticated algorithms take longer for complex functions than for simple ones.  For example, MARS required a far longer runtime for the sinusoid with 10,000 points, but also produced a much better fit as a result.  As a general rule, algorithms with shorter runtimes have higher bias and variance, with running medians both quickest and producing the worst fit.}
\end{itemize}

The FINEST algorithm for any particular problem depends upon which of these criteria are most critical.  For typical astronomical problems, it is likely that fit quality, safety, and interpretability will be most important.  Running averages and medians perform worst in each of these, and it seems clear that these algorithms should be avoided under almost all circumstances given the availability of easy and fast alternatives.  Although running medians are simplest and fastest, their advantages in both categories are minor over most other algorithms and outweighed by the poor performance of smoothing algorithms when used for modeling problems.  For rare, extremely computationally limited cases, it may be the only algorithm that converges in available processing time, and this is the only circumstance under which the authors would recommend its usage.

\subsection{Two-Error Regressions}

This work also presents \actwod, a new algorithm that can be used for a general 2-E nonparametric regression.  As described in \S~\ref{sec:2d}, it will produce a fit for an arbitrary relationship convolved with a two-error error function.  We recommend using \actwod\ for two-error nonparametric regressions simply because of the lack of alternatives.  However, two-error fully nonparametric regression is very difficult, and in most cases a better solution will involve additional assumptions about the underlying relationship between $X$ and $Y$.  For example, the knowledge that the underlying relationship is expected to be monotonic will typically produce a much better fit.  

The one-error problem is a subset of the two-error problem for the case in which errors in one variable are negligible.  Thus, it is natural to consider whether \actwod\  or other two-error methods should be applied in all cases.  The problem is that the two-error problem is far more difficult, and thus the solution is both more computationally expensive and more approximate.  \ac\ and Kriging are formally minimum variance estimators for the one-error problem.  Gaussian deconvolution is in some sense a two-error analogue of Kriging.  However, the process of turning a deconvolution into a predictor in two dimensions is neither formally nor practically minimum variance.  Thus, except for the case in which the entire dataset is well described by one Gaussian, one-error methods will produce both a quicker and better fit for one-error problems than two-error methods will.  As with nearly every regression problem, nonparametric methods are a blunt instrument, and with additional information, a more appropriate tool will yield a better result.

\subsection{Does It Really Matter?}

For many of the examples given in \S~\ref{sec:results}, even running medians and running averages produced apparently reasonable fits.  They are still dangerous because of the possibility of silent, catastrophic failure.  However, most datasets are not pathological.  Even for those cases, however, it should be noted that other regression algorithms provide a better predictor.  In effect, using running medians rather than, e.g., Kriging, means that more data is needed in order to produce a predictor of the same accuracy and precision.  

\begin{figure}[h]
	\plotone{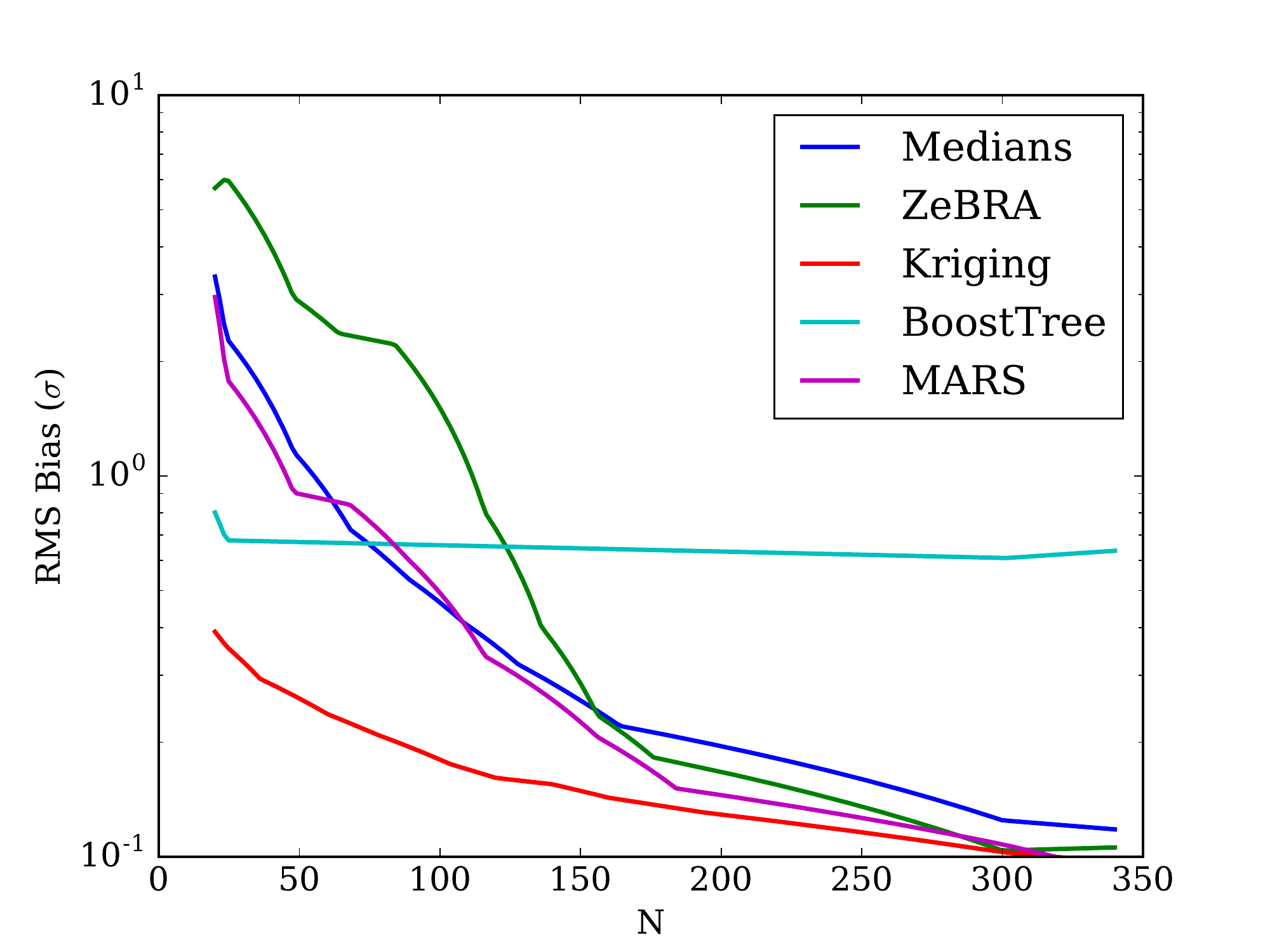}
	\caption{The typical error in the estimated depth of the exoplanet transit light curves from Fig. \ref{fig:lightcurve} is shown for several of the 1D methods we have discussed. A running bin of five items was used for computing medians. The curves were determined by sampling each method $400$ times at each of several $N$ and then fitting the resulting point cloud using \ac, which was chosen according to the flowchart (Fig. \ref{fig:flowchart}) because the data was well-sampled, we did not need an analytically interpretable result, and we wanted a low-bias fit.}
	\label{fig:findN}
\end{figure}

As an example, we return to the mock exoplanet transit light curve fit earlier and consider another common problem, finding the depth of the eclipse in order to calculate the ratio of the exoplanet radius to the star's radius, and thus ultimately the radius of the exoplanet\footnote{Although we have chosen this as an example because it is easy to demonstrate where the improvement is coming from, exoplanet transit analyses typically make assumptions about the shape of the exoplanet light curve rather than using nonparametric regression.  When such assumptions are correct, the additional information they provide will generally produce a superior fit using optimal techniques, which is why parametric methods are a better choice for well-modeled problems.}.  Algorithms producing better predictors will, on average, require fewer points to measure the depth to the same precision.  The uncertainty in measuring the depth as a function of the number of points available is highly dependent upon the choice of algorithm (Fig. \ref{fig:findN}).

For large $N$, one of the problems with running averages and medians is that the bias ends up producing a minimum possible variance.  However, even for the case of a Gaussian distribution, for which averages and medians are unbiased, Boosted Trees and Kriging still require considerably fewer points to determine the depth of a transiting light curve to within $1.0 \sigma$ (the size of the noise added), roughly $20$ points versus $50$. At $0.3\sigma$ Kriging performs vastly better, still requiring only $20$ points versus $120$ for medians and MARS and $140$ for \ac. The strong performance of medians in this case is primarily because the depth is an integrated measure, so that for this particular metric the much higher point-to-point variance (Fig. \ref{fig:lightcurve2}) is essentially smoothed over the entire transit, with the error dominated by bias.  The weak performance of Boosted Trees at larger $N$ is similarly due to its preference for low variance over low bias. This merely reinforces the point: the proper choice of regression technique depends heavily upon the problem at hand.  Even in this case, Boosted Trees substantially outperforms, e.g., \ac~for small $N$ but the reverse is true at large $N$.

It may be counterintuitive that even in a situation when averaging is unbiased, it still produce suboptimal performance.  However, a similar effect occurs in many problems relating to unbiased estimators, most notably the ``German Tanks'' problem \citep{Ruggles1947}.  Even though the average will converge to the right answer, averaging the dataset is not the best way of estimating the average value that would result from taking a new observation from an identical sample.  And although in some cases the improvement from superior techniques may be slight, there is no more justification for choosing a poor approximation than for continuing to use the trapezoidal rule rather than using more advanced numerical integration methods that are already in existing libraries and just as simple to run.  

In summary, because there are tradeoffs in choosing numerical methods for nonparametric regression, we do not advocate for any particular method.  We do, however, believe that the landscape of these tradeoffs has changed over time, and that some of the weakest methods no longer have a place in the scientific repertoire.  The often-implicit decision to optimize for computational resources which prevailed in the past century played a valuable role in enabling scientists to come to grips with their data.  In the 21st century, however, this decision is no longer a requirement, and as we have shown, continuing to prioritize simplicity of method and computational speed over statistical robustness is a dangerous path.

The advent of increasingly expensive, large telescopes makes using the best statistical techniques available even more essential.  The most expensive telescopes now have operating costs of several dollars per second.  If the same science can be done with improved statistical techniques, that expensive observing time would be replaced with far cheaper effort on the part of astronomers ($\sim$ \$1 per minute) or computers ($\sim$ \$1 per day).  

Just as using an improved technique for the exoplanet transit allows the radius to be measured to higher precision, equivalently it means that the same science can be performed with a factor of $>6$ less observing time.  Indeed, if every problem scaled similarly, the reduced time requirements gained by using improved statistical methods would allow, e.g., the Hubble Space Telescope to award time to the top three quintiles of submitted proposals.

The tools we develop today will be the legacy code of tomorrow, and may be applied well outside of the domain we expect, by scientists who are not fully familiar with their design parameters.  Furthermore the datasets of interest in astronomy are becoming increasingly precise, and we are coming to ask increasingly subtle questions of them.  Without knowing where bias may lie, we may easily be led astray, so there is no longer justification for choosing frequently-biased methods that can fail catastrophically and silently over better techniques that are safer, more interpretable, and produce a better fit.

\acknowledgments

The authors would like to thank Richard Yi for several helpful conversations in developing these ideas and the anonymous referee for a review process that substantially improved this work.  The authors would also like to thank Jogesh Babu, Douglas Boubert, Peter Capak, Jens Hjorth, Nick Lee, Dan Masters, Josh Speagle, and Sune Toft for helpful comments.  CS acknowledges support from the ERC Consolidator Grant funding scheme (project ConTExt, grant number No. 648179) and from the Carlsberg Foundation.  The Cosmic Dawn Center is funded by the Danish National Research Foundation.  ASJ is supported by a Marshall scholarship.

\end{document}